\documentclass{aa}

\usepackage[T1]{fontenc}

\DeclareRobustCommand{\VAN}[3]{#2}
\let\VANthebibliography\thebibliography
\def\thebibliography{\DeclareRobustCommand{\VAN}[3]{##3}\VANthebibliography}

\usepackage{graphicx}	
\usepackage{subcaption}
\usepackage{placeins}
\usepackage{siunitx}	
\usepackage{amsmath}	
\usepackage{amssymb}	
\usepackage{aas_macros}
\usepackage{enumitem} 
\usepackage[shortcuts,acronyms]{glossaries-extra}		
\usepackage{orcidlink}
\usepackage{amsfonts}
\usepackage{float}
\usepackage{placeins}

\begin{document}

\title{The tidal features of the classical Milky Way satellites: Expected in MOND but inconsistent with cold dark matter models}

\author{
Elena Asencio\orcidlink{0000-0002-3951-8718}$^{1}$\thanks{E-mail: s6elena@uni-bonn.de (EA)},
Indranil Banik\orcidlink{0000-0002-4123-7325}$^{2}$
and Pavel Kroupa\orcidlink{0000-0002-7301-3377}$^{1,3}$}

\institute{
$^{1}$Helmholtz Institut für Strahlen und Kernphysik, Universität Bonn, Nussallee 14-16, 53115 Bonn, Germany\\
$^{2}$Institute of Cosmology \& Gravitation, University of Portsmouth, Dennis Sciama Building, Burnaby Road, Portsmouth PO1 3FX, UK \\
$^{3}$ Astronomical Institute of Charles University, Faculty of Mathematics and Physics, V Hole\v{s}ovi\v{c}kách 2, Praha, CZ-180~00, Czech Republic\\
}

\date{Accepted XXX. Received YYY; in original form ZZZ}

\abstract
{Most classical satellites of the Milky Way are known to display signs of tidal disturbance (e.g. tidal tails, substructures, and distorted shapes). This cannot be explained by the standard model of cosmology due to its prediction that the dark matter haloes of the classical satellites confer them with very strong self-gravity and make them resilient to the Milky Way's gravitational tides. In this work, we estimate the tidal susceptibility of the classical satellites by comparing their half-mass radius with their theoretical tidal radius at pericentre in both the standard model and in the Milgromian dynamics (MOND) model. With this approach, we demonstrate that most classical satellites are expected to be tidally perturbed in MOND, so their observed tidal features are generally in good agreement with MOND expectations. Since gravitational tides can also enhance the velocity dispersion of the satellites, we argue that MOND can plausibly explain the unusually high velocity dispersions reported for some of the classical satellites.}


\titlerunning{The tidal features of the classical satellites}
\authorrunning{E. Asencio, I. Banik \& P. Kroupa}

\maketitle
\nolinenumbers

\glsresetall

\section{Introduction}
\label{Introduction}

The Milky Way (MW) is surrounded by several dwarf galaxies, many of which orbit around its gravitational potential. These galaxies are known as MW satellite galaxies. Observations of them have been used in several studies to understand their properties as well as their formation histories and incorporation into the MW potential. Notably, the implications of these studies are also relevant for testing cosmological and gravitational models that make specific predictions on the structure and formation of galaxies.

For instance, according to the Lambda cold dark matter ($\Lambda$CDM) model \citep{Efstathiou_1990, Ostriker_1995} $-$ the current standard cosmological model $-$ most galaxies are embedded in dark matter (DM) haloes and form through mergers of smaller structures, such as  dwarf galaxies. The model also predicts that galaxies should still be surrounded by several dwarf galaxies distributed in an approximately isotropic way \citep{Moore_1999, Gao_2004}. However, the planar distribution of the satellite galaxies observed around the MW, M31, and Centaurus A contradicts this prediction \citep{Lynden_Bell_1976, Kroupa_2005, Metz_2007, Metz_2008, Metz_2009, Ibata_2013, Ibata_2014, Tully_2015_Cen_A, Muller_2018, Pawlowski_2021}.

Because of this contradiction, an alternative formation scenario has been proposed in the literature \citep{Metz_Kroupa_2007, Pawlowski_2012, Zhao_2013, Banik_2018, Bilek_2018, Banik_2022}. In this scenario, the satellite galaxies formed in the tidal tails resulting from an intergalactic interaction, making them tidal dwarf galaxies (TDGs). This seems to be a more natural explanation since structures that originate from the same tidal tail are expected to be planar and coherent in their motion \citep{Pawlowski_2011, Kroupa_2012, Pawlowski_2018, Haslbauer_2019, Bilek_2021}. Although TDGs can form in a $\Lambda$CDM framework, they are DM-free in this paradigm \citep{Barnes_1992, Braine_2001, Wetzstein_2007} because the potential wells of the tidal tails are not deep enough to retain DM particles due to their high velocity dispersion. This implies that the baryonic mass of a TDG should be approximately the same as the mass inferred from its internal dynamics assuming virial equilibrium and Newtonian gravity. However, this is not the case for the satellite galaxies, where velocity dispersions exceed Newtonian expectations \citep{Armandroff_1995, Burkert_1997, Tollerund_2013, McGaugh_2013a}. Several works \citep{Kroupa_1997, Metz_2007, Casas_2012} have proposed that the high velocity dispersions are not caused by the presence of DM haloes but by tidal interactions of the satellite galaxies with the main galaxy. This would also explain why some of them appear to be highly perturbed \citep{Kroupa_2010}. However, in this scenario, the satellite galaxies would need to be observed at the right time, along the right orbit, and during the quasi-equilibrium phase of their evolution in order to match the observed velocity dispersions. Otherwise, their velocity dispersions would be far too high \citep{Metz_2007, Casas_2012}. Moreover, TDGs are expected to be very fragile and short-lived in $\Lambda$CDM, as they lack the self-gravity `boost' provided by DM haloes. Therefore, it would be extremely unlikely that so many of them would still exist in the Local Group \citep{Bournaud_2006, Haslbauer_2019}.

The elevated internal velocity dispersions are generally not a problem for alternative models that do not resort to DM for explaining the structure and dynamics of galaxies. The most successful among these models is currently Milgromian dynamics \citep[MOND;][]{Milgrom_1983}. In MOND, gravity experiences a boost in the regime of low accelerations, i.e. when the gravitational acceleration ($g$) is $g~\la~a_0 = 1.2 \times 10^{-10}~\textrm{m/s}^2$ \citep{Begeman_1991, Gentile_2011}. Therefore, in this model, the enhanced gravitational acceleration inferred on galactic scales (especially in dwarf galaxies) is attributed to a non-Newtonian behaviour of gravity rather than to an unobservable field of particles that could potentially be separated from the observable galaxy. This means that MOND can simultaneously explain the high internal velocity dispersions of the Local Group satellites and their planar distribution \citep[for a comprehensive review of the MOND model, see][]{Famaey_McGaugh_2012, Banik_Zhao_2022}.

Interestingly, the planar distribution of the satellite galaxies is not the only dwarf galaxy-related observation for which MOND provides a more successful interpretation. \citet{Asencio_2022} found that for the Fornax galaxy cluster, the fraction of morphologically disturbed dwarf galaxies and their distribution is highly consistent with MOND expectations but inconsistent with expectations based on the $\Lambda$CDM model. This is because DM haloes provide the dwarfs with too much protection against cluster tides, while MOND manages to strike the right balance between allowing the dwarfs to be morphologically disturbed and providing some degree of protection against tides. This is due to the combination of the MONDian enhancement to Newtonian gravity and the external field effect \citep[EFE;][]{Milgrom_1986} that arises from the non-linearity of MOND.

Several works have also reported signs of tidal disturbance in the morphology and other properties of several MW satellite galaxies as well as other properties \citep{Hodge_1964, Hodge_Michie_1969, Lynden-Bell_1982, Irwin_1995, Kroupa_1997, Eskridge_2001, Munoz_2005, McGaugh_Wolf_2010, Battaglia_2012, Sestito_2023, Bilek_2025}. These include such signatures as enhanced velocity dispersions, episodic star formation that coincides with pericentre passages, large ellipticities (or variations of the dwarf ellipticity with radius), asymmetries in the density distribution of the dwarf, unbound stellar populations near the dwarf, and tidal tails. Similar to the Fornax Cluster dwarfs, these signatures cannot generally be ascribed to the effect of tides in $\Lambda$CDM due to its prediction that the dwarfs are surrounded by a protective DM halo \citep{Penarrubia_2009, Battaglia_2015, Iorio_2019, Boyea_2026}.

In this work, we analyse the tidal susceptibility of the MW satellite galaxies in MOND to check if this framework is capable of providing a better explanation for their observed characteristics. In particular, we focus on the classical satellites \citep{Pawlowski_2020}, as they are some of the best observed and most thoroughly studied satellites of the MW.

Our paper is organised as follows. In Section~\ref{Classical_sat} we describe the observational data used in our analysis. In Section~\ref{rtid_eta} we explain how we obtained the theoretical tidal radius and the tidal susceptibility of the satellites in MOND and in $\Lambda$CDM (Section~\ref{rtid_eta_theo}) and how we calculated the value of both parameters in each theory for each of the considered satellite galaxies (Section~\ref{rtid_eta_sat}). In Section~\ref{N-body} we describe the features expected in tidally disturbed dwarf galaxies according to $N$-body simulations. In Section~\ref{Observations} we compare the tidal features predicted by $N$-body simulations with the observed features of the classical satellites. In Section~\ref{Discussion} we discuss several aspects of our results: We further describe the individual tidal features of the classical satellites provided by previous studies (Section~\ref{Discussion_satellites}), we discuss whether a different DM mass fraction could reconcile the observations of tidal features in the classical dwarfs with CDM models (Section~\ref{DM_mass}), and we address claims from previous works regarding the high velocity dispersions of some of the classical satellites (Section~\ref{Other_studies}). We conclude in Section~\ref{Conclusions}. Appendices~\ref{MOND_Nbody_sim}, \ref{rtid_eta_present}, and \ref{images_classical_sat} provide additional technical material and a collation of the isophotal appearence of the classical satellite galaxies.

\section{The classical MW satellites: Observational data and inferred orbital parameters}
\label{Classical_sat}
Sagittarius, the Large Magellanic Cloud, the Small Magellanic Cloud, Draco, Ursa Minor, Sculptor, Sextans, Carina, Fornax, Leo~I, and Leo~II are considered to be the 11 classical satellites of the MW. This denomination comes from the fact that these were some of the first stellar objects to be observed and identified as MW satellite galaxies \citep{Shapley_1939, Hodge_1971}. Their favourable visibility has enabled several detailed studies on their morphology and features \citep{Hodge_1963, Irwin_1995, Piatek_2001, Westfall_2006, Coleman_2007, Battaglia_2011} as well as fairly precise measurements of their position, kinematics, stellar mass, and more recently their proper motions \citep{Lokas_2009, Pace_2022, Bennet_2024}.

In this work, we use this information to infer the orbital parameters of the classical satellites (Table~\ref{orbital_prop}), as well as their expected tidal susceptibility (Section~\ref{rtid_eta}). Since the main goal of our study is to analyse the tidal effects of the MW on the dwarf properties, we remove Sagittarius and the two Magellanic clouds from our sample. The reason for this is that Sagittarius is thought to have been affected by interactions with the MW disk \citep{Ibata_1998, Laporte_2019}, while the Magellanic clouds are disrupting each other already \citep{Mackey_2016, De-Leo_2020}.

For our analysis, we used the parameter values (heliocentric distance, RA, Dec, radial velocity, sky-projected half-mass radius, absolute $V$-band magnitude, and proper motions) of the classical satellites as presented in tables~1 and 2 of \citet{Pace_2022}. Since the more recent study of \citet{Bennet_2024} provides more accurate proper motions for Leo~I and Leo~II, we used the proper motions of that study instead for these specific satellites. For the internal line of sight velocity dispersion ($\sigma_v$) values, we used the data from \citet{McConnachie_2020}.

We converted the Heliocentric quantities given in these studies to a Galactocentric frame assuming a Sun to Galactic centre distance of 8.2~kpc, with the circular speed at the Sun being 232.8~km/s \citep{McMillan_2017}. For the solar motion with respect to the local standard of rest (LSR), we assumed (U, V, W) = (14.1, 14.6, 6.9)~km/s \citep{Francis_Anderson_2014}.

Then, using the fourth-order Runge-Kutta method for orbital integrations, we computed the Galactocentric positions and velocities of the satellites throughout their orbit around the MW. In order to account for the uncertainties in the observational data (mainly the distance and proper motions), we constructed a grid of orbital integrations that samples the uncertainties in these parameters.

For the MW potential in our analytical simulation, we assumed a point mass of $M_{\rm b, MW} = 6.7 \times 10^{10}~M_{\odot}$ \citep[MW's disk baryonic mass; ][]{Banik_2018_escape} and obtained its total gravitational acceleration by applying the MOND equation: 
\begin{eqnarray}
	g ~=~ g_{_N} \nu \left(g_{_N}\right) \, ,
	\label{g_g_N}
\end{eqnarray}
where $\nu \left(g_{_N}\right)$ in our analysis is the ``simple'' interpolating function proposed by \citet{Famaey_Binney_2005}:
\begin{eqnarray}
	\nu \left( g_{_N} \right) ~=~ \frac{1}{2} + \sqrt{\frac{1}{4} + \frac{a_{_0}}{g_{_N}}} \, .
	\label{simple_interpolating}
\end{eqnarray}
This function depends on the Newtonian gravity $g_{_N} = G M_{b}/D^2$, with $G$ being Newton's constant, $M_b$ the baryonic content of the Galaxy, and $D$ the Galactocentric distance. This particular formulation of MOND is known as the ``quasi-linear formulation of MOND'' \citep[QUMOND,][]{Milgrom_2010}. For the purpose of the present calculation, the MONDian boost to $g_N$ is equivalent to having a DM halo contributing to the MW potential. Therefore, we used this MW $g$ value both for the MOND and the $\Lambda$CDM analyses. We refer to $g$ estimated in this way as
\begin{eqnarray}
	g_{\textrm{MW}} =  \underbrace{g_{N, MW} \, \nu \left(g_{N, MW}\right)}_{\textrm{MOND}} \equiv \underbrace{g_{N, MW} + \frac{G M_{\textrm{DM, MW}}}{D^2}}_{\Lambda\textrm{CDM}} \, ,
\end{eqnarray}
where the contribution of the DM term can be inferred under the assumption that the $\Lambda$CDM value of $g_{\textrm{MW}}$ matches the predicted MONDian $g_{\textrm{MW}}$ at every $D$.\footnote{The assumption that the Galactic potential is well approximated by MOND follows from its success with galaxy rotation curves \citep{Famaey_McGaugh_2012, Banik_Zhao_2022}.} This gives us $M_{\textrm{DM, MW}} \approx 7.2 \times 10^{11}~M_{\odot}$ at $D \approx 100$~kpc, the distance at which we find most of the considered satellite galaxies. This value is similar to Newtonian dynamical estimates of the MW mass at that distance \citep{Deason_2021}. For consistency, in our following analyses, we use the MONDian form of $g_{\textrm{MW}}$ for performing calculations within the MOND framework (e.g. Eq.~\ref{rtid_MOND}) and the $\Lambda$CDM form of $g_{\textrm{MW}}$ for $\Lambda$CDM calculations (e.g. Eq.~\ref{rtid_LCDM}). Table~\ref{internal_prop} shows the MW satellite internal properties relevant for this study, while Table~\ref{orbital_prop} shows the orbital properties of the satellites inferred through our orbit integration method.

\begin{table*}
	\caption{Internal properties of the classical satellites.}
	\centering
	\begin{tabular}{c|c|c|c|c|c}
		 \hline
		Name     & $M_{\star}$ ($10^6 M_{\odot}$)     & $r_{h, \mathrm{3D}}$ (pc)    &    $\epsilon$    &  $\sigma_v$ (km/s)     & $t_{\textrm{cross}}$ (Myr)        \\ \hline
		\multicolumn{1}{c|}{Fornax}     & $40.27^{+5.54}_{-4.87}$ & $1111.78 \pm 3.35$ & $0.29 \pm 0.02$ &  $11.7 \pm 0.9$ & $185.85 \pm 14.31$ \\ 
		\multicolumn{1}{c|}{Carina}     & $0.98^{+0.05}_{-0.04}$ & $404.80 \pm 4.01$ & $0.36 \pm 0.01$  &  $6.6 \pm 1.2$ & $119.96 \pm 21.84$ \\ 
		\multicolumn{1}{c|}{Draco}      & $0.51^{+0.02}_{-0.02}$ & $278.20 \pm 2.59$ & $0.29 \pm 0.01$ & $9.1 \pm 1.2$ & $59.79 \pm 7.90$   \\ 
		\multicolumn{1}{c|}{Ursa Minor} & $0.68^{+0.03}_{-0.03}$ & $529.25 \pm 3.18$ & $0.55 \pm 0.01$ & $9.5 \pm 1.2$ &  $108.96 \pm 13.78$  \\ 
		\multicolumn{1}{c|}{Leo~I}      & $8.57^{+2.52}_{-1.95}$ & $357.69 \pm 2.94$ & $0.3 \pm 0.1$ & $9.2 \pm 1.4$ & $76.04 \pm 11.59$    \\ 
		\multicolumn{1}{c|}{Leo~II}     & $1.31^{+0.05}_{-0.05}$ & $222.85 \pm 2.65$ & $0.07 \pm 0.01$ & $6.6 \pm 0.7$ & $66.04 \pm 7.05$    \\ 
		\multicolumn{1}{c|}{Sculptor}   & $3.54^{+0.49}_{-0.43}$ & $355.69 \pm 1.59$ & $0.33 \pm 0.01$ & $9.2 \pm 1.4$ & $75.62 \pm 11.51$   \\ 
		\multicolumn{1}{c|}{Sextans}    & $0.51^{+0.03}_{-0.03}$ & $579.27 \pm 3.51$ & $0.30 \pm 0.01$ & $7.9 \pm 1.3$ & $143.41 \pm 23.62$   \\ \hline
	\end{tabular}
	\tablefoot{The columns correspond to the following quantities: (1) names of the classical satellites; (2) stellar mass ($M_{\star}$) obtained from the observed absolute $V$-band magnitude ($M_V$) in \citet{Pace_2022} for a mass to light ratio of 2 \citep{Telford_2020}; (3) 3D half-mass radius ($r_{h, \mathrm{3D}}$) obtained from the projected 2D half mass radius ($r_h$) in \citet{Pace_2022} and the assumption that the internal density profile of the satellites follows a Plummer profile, $r_{h, 3D} = r_h / \sqrt{4^{1/3} - 1}$ \citep{Plummer_1911, Binney_Tremaine}; (4) ellipticity ($\epsilon$) of the satellites taken from \citet{Pace_2022}; (5) line of sight internal velocity dispersion of the satellites ($\sigma_v$) taken from \citet{McConnachie_2020}; (6) crossing time ($t_{\textrm{cross}} \equiv 2 \, r_{h, 3D}/\sigma_v$) \citep{Binney_Tremaine}.}
	\label{internal_prop}
\end{table*}

\begin{table*}
	\caption{Orbital properties of the classical satellites.}
	\centering
	\begin{tabular}{c|c|c|c|c|c}
		\hline		
		Name     & $D_{\textrm{obs}}$ (kpc)     & $D_{\textrm{peri}}$ (kpc)    &    $e$    &  $t_{\textrm{orb}}$ (Gyr)    & $t_{\textrm{peri}}$ (Gyr)        \\ \hline
		\multicolumn{1}{c|}{Fornax}     & $149.2 \pm 8.4$ & $76.41^{+31.73}_{-23.69}$ & $0.23^{+0.06}_{-0.01}$ & $2.72^{+0.53}_{-0.37}$ & $1.67^{+0.43}_{-0.31}$ \\ 
		\multicolumn{1}{c|}{Carina}     & $107.2 \pm 5.4$ & $108.14^{+8.28}_{-11.54}$ & $0.04^{+0.10}_{-0.01}$  &  $2.46^{+0.45}_{-0.31}$ & $1.23^{+0.31}_{-0.21}$ \\ 
		\multicolumn{1}{c|}{Draco}      & $75.8 \pm 5.4$ & $44.82^{+5.47}_{-4.82}$ & $0.49^{+0.02}_{-0.02}$ & $1.70^{+0.17}_{-0.16}$ & $1.32^{+0.14}_{-0.15}$   \\ 
		\multicolumn{1}{c|}{Ursa Minor} & $78.1 \pm 4.2$ & $45.57^{+4.13}_{-3.36}$ & $0.45^{+0.03}_{-0.02}$ & $1.64^{+0.11}_{-0.09}$ & $1.21^{+0.08}_{-0.08}$  \\ 
		\multicolumn{1}{c|}{Leo~I}      & $262.0 \pm 9.5$ & $52.44^{+4.67}_{-4.81}$ & $0.91^{+0.01}_{-0.01}$ & $5.90^{+0.18}_{-0.16}$ & $1.01^{+0.03}_{-0.04}$     \\ 
		\multicolumn{1}{c|}{Leo~II}     & $235.6 \pm 15.0$ & $77.55^{+13.17}_{-11.31}$ & $0.66^{+0.06}_{-0.07}$ & $3.78^{+0.29}_{-0.26}$ & $1.70^{+0.12}_{-0.11}$    \\
		\multicolumn{1}{c|}{Sculptor}   & $84.0 \pm 1.5$ & $59.92^{+2.99}_{-3.63}$ & $0.39^{+0.02}_{-0.01}$ & $1.97^{+0.09}_{-0.08}$ & $0.42^{+0.02}_{-0.03}$   \\
		\multicolumn{1}{c|}{Sextans}    & $95.5 \pm 2.5$ & $83.66^{+3.30}_{-3.19}$ & $0.47^{+0.04}_{-0.03}$ & $3.03^{+0.25}_{-0.23}$ & $0.26^{+0.03}_{-0.04}$   \\ \hline
	\end{tabular}
	\tablefoot{The columns correspond to the following quantities: (1) name of the satellite, (2) Galactocentric distance (D) at currently observed position, (3) Galctocentric distance at pericentre ($D_{\textrm{peri}}$), (4) eccentricity of the orbit ($e$), (5) time required to complete a full orbit ($t_{\textrm{orb}}$), (6) time since last pericentric passage ($t_{\textrm{peri}}$).}
	\label{orbital_prop}
\end{table*}

\section{Tidal radius and tidal susceptibility}
\label{rtid_eta}
\subsection{Relevant equations}
\label{rtid_eta_theo}
In order to estimate the susceptibility of the classical satellites to the tidal force of the MW, we use the method described in sections 3 and 4 of \citet{Asencio_2022}. In other words, we compare the tidal radius ($r_{\textrm{tid}}$) of each satellite with respect to its current deprojected half-mass radius $r_{h, \mathrm{3D}}$ (see Table~\ref{internal_prop}). The ratio of these two quantities is the tidal susceptibility ($\eta$) of the satellite:
\begin{eqnarray}
	\eta \equiv \frac{r_{h, \mathrm{3D}}}{r_{\textrm{tid}}} \, ,
	\label{eta}
\end{eqnarray}
where $r_{\textrm{tid}}$ is the radial distance, measured from the centre of the satellite galaxy, beyond which gravitational tides from the main galaxy dominate over the self-gravity of the satellite galaxy. Therefore, satellite galaxies with $r_{\textrm{tid}} \ll r_{h, \mathrm{3D}}$ are expected to be very disturbed (or even destroyed) by tides, while satellites with $r_{\textrm{tid}} \gg r_{h, \mathrm{3D}}$ would be stable. Since we are interested in the maximum $\eta$ experienced by the satellite throughout its orbit, we measure this quantity when the satellite is at perigalacticon for all our nominal results.

For the $\Lambda$CDM analysis, we used the $r_{\textrm{tid}}$ expression obtained from eq.~1 of \citet{Baumgardt_2003} under the assumption of Newtonian gravity:
\begin{eqnarray}
	r_{\textrm{tid, Newton}} ~=~ \left( \frac{G \, M_{\textrm{dwarf, tot}}}{ \mathrm{d} g_{_{\textrm{MW}}} /  \mathrm{d} D} \right)^{1/3} \, ,
	\label{rtid_LCDM}
\end{eqnarray}
where $ \mathrm{d} g_{_{\textrm{MW}}} /  \mathrm{d} D$ is the tidal stress from the MW and $M_{\textrm{dwarf, tot}}$ is the total mass of the dwarf. If the dwarf is assumed to be a Newtonian TDG, $M_{\textrm{dwarf, tot}}$ will only be composed of the baryonic mass of the dwarf (for this scenario, we refer to Eq.~\ref{rtid_LCDM} as $r_{\textrm{tid, TDG}}$). Otherwise, its $M_{\textrm{dwarf, tot}}$ is obtained from its known baryonic mass $-$ that is, its stellar mass ($M_{\star}$) $-$ and its DM mass. Since DM cannot be observed, the DM fraction of a dwarf is inferred either from $\Lambda\textrm{CDM}$ simulations or from the dwarf's internal dynamics. Fig.~16 in \citet{Asencio_2022} shows a comparison between DM fractions obtained from the simulations of \citet{Moster_2010}, the cosmological simulation Illustris TNG50 \citep{Pillepich_2018, Pillepich_2019, Nelson_2019b, Nelson_2019} and the dynamics of isolated dwarf galaxies of the Local Group \citep{Falcon_Barroso_2011, McConnachie_2012, Rys_2014, Toloba_2014}. The DM fraction obtained from simulations constitutes the DM fraction predicted by the $\Lambda$CDM model, while the DM fraction obtained from dynamics corresponds to an observational fit. For this study, we choose the dynamical mass method to obtain the DM fraction of the dwarf galaxies for various reasons: it is quite possibly the most accurate description of the Local Group dwarf galaxies within this framework, as it is calibrated with observations; it constitutes one of the most conservative DM fractions to assume within a $\Lambda$CDM framework, especially at the low-mass end \citep[see fig.~16 of][]{Asencio_2022}; and, since it does not depend on predictions made specifically by the $\Lambda$CDM model, it serves as a more general test for CDM models. Because of this, in the following, we use the term CDM instead of $\Lambda$CDM for our nominal standard framework analysis, to signify that our results apply to CDM models in general, not only $\Lambda$CDM. 

The dynamical mass of a galaxy within its (baryonic) $r_{h, \mathrm{3D}}$ is given by eq.~2 in \citet{Wolf_2010}:
\begin{eqnarray}
    M_{\textrm{dyn}} \left( < r_{h, \mathrm{3D}} \right) ~=~ \frac{3 \, r_{h, \mathrm{3D}} \langle \sigma^2_{v} \rangle}{G} \, .
    \label{M_dyn}
\end{eqnarray}
Galactic tides can affect the $\sigma_v$ of a dwarf galaxy \citep{Piatek_1995, Kroupa_1997, Read_2006, McGaugh_Wolf_2010} and, therefore, its estimated $M_{\textrm{dyn}}$. Because of this, we do not infer the DM content of the classical satellites directly from their $\sigma_v$. Instead, we used the relation between $M_{\star}$ and $M_{\textrm{dyn}}$ obtained in fig. 16 of \citet[][]{Asencio_2022} for a group of isolated nearby dwarf galaxies. This led to the following $M_{\star}-M_{\textrm{DM}}$ relation \citep[eq.~1 in][]{Asencio_2022_erratum}\footnote{We note that the original \citet{Asencio_2022} paper had a typo in the text regarding the equation relating $M_{\textrm{DM}}$ and $M_{\star}$ (eq.~65), so we point to the corrected version in the \citet{Asencio_2022_erratum} erratum. The typo did not affect the calculations in \citet{Asencio_2022}.}:
\begin{eqnarray}
     \log_{10} \left( \frac{M_{\textrm{DM}}}{M_{\star}} + 1 \right) ~=~ 4.089 - 0.396~\log_{10} \left( \frac{M_{\star}}{M_{\odot}} \right) \, .
    \label{DM_frac}
\end{eqnarray}
Therefore, our nominal total dwarf mass ($M_{\textrm{tot, CDM}}$) for the CDM paradigm was obtained under the assumption of a Newtonian gravitational law and of a DM mass fraction consistent with that of isolated dwarf galaxies in the Local Group (we refer to Eq.~\ref{rtid_LCDM} obtained for $M_{\textrm{tot, CDM}}$ as $r_{\textrm{tid, CDM}}$). For comparison, we also repeated our CDM analysis for a DM mass fraction obtained with Eq.~\ref{M_dyn} and the classical satellite's $\sigma_v$ (we refer to the $r_{\rm tid}$ obtained with this method as $r_{\textrm{tid, dyn}}$) and for the scenario in which the classical dwarfs are TDGs $-$ that is, DM-free. We discuss these last two cases in Section~\ref{DM_mass}.

The $r_{\textrm{tid}}$ corresponding to the MOND framework ($r_{\textrm{tid, MOND}}$) differs from Eq.~\ref{rtid_LCDM}, not only because of the absence of a DM component in this model, but also due to the MONDian modification to Newton's gravity law (see Eqs.~\ref{g_g_N} and ~\ref{simple_interpolating}). The non-linearity of MOND leads to a breach of the strong equivalence principle. That is, in MOND, the internal dynamics of galaxies can be affected by the gravitational field of external structures. This effect is known as the EFE. An EFE-dominated satellite experiences a reduction in its self-gravity's MONDian boost, making the satellite more susceptible to tides. This imples that the satellite galaxies can be affected by the external field of the MW and are therefore expected to have lower $r_{\textrm{tid}}$ values (and higher $\eta$ values) than in the CDM model, given the self-gravity is similar in both theories for isolated galaxies.

The derivation of the MOND tidal radius is described in section~3.3 of \citet{Asencio_2022} and is based on eqs.~26 and 36 of \citet{Zhao_2006}:
\begin{eqnarray}
	\label{rtid_MOND}
	&& r_{\textrm{tid, MOND}} \\
	&& = \frac{2}{3} \left. \sqrt{ \frac{\partial \ln g}{\partial \ln g_{_N}}} \right|_{g = g_{_{\textrm{MW}}}} \left[ \left( \frac{2 - \alpha}{3 - \alpha} \right) \frac{G_{\textrm{eff}} \, M_{\star}}{d g_{_{\textrm{MW}}} / d D} \right]^{1/3}, \nonumber
\end{eqnarray}
with $\alpha \equiv 2 + (d\ln g_{_\textrm{MW}} / d\ln D)$. The term $G_{\textrm{eff}} ~=~ G(a_{_0} + g_{_{MW}})/g_{_{MW}}$ \citep[eq.~19 in][]{Asencio_2022} is the MONDian correction to Newton's $G$ for the simple interpolating function, derived under the assumption that $g_{_{MW}}$ dominates over the satellite's gravity.

Especially in the MOND case, we note that the satellites which are significantly affected by tides and/or experience different gravitational regimes throughout their trajectories (e.g. satellites which are considerably affected by the EFE with relatively high orbital eccentricities, such as Draco, Ursa Minor, Sextans and $-$ to a smaller extent $-$ Sculptor) are expected to have their $r_{h, \mathrm{3D}}$ values notably expanded after their pericentric passages \citep{Brada_2000_tides, Asencio_2022}. Since Sculptor and Sextans are currently observed shortly after their pericentric passage, it is possible that their observed $r_{h, \mathrm{3D}}$ is an overestimation of their $r_{h, \mathrm{3D}}$ at pericentre. Assuming that the degree of expansion of $r_{h, \mathrm{3D}}$ shortly after pericentre in the classical MW satellites is similar to that of the Fornax dwarfs \citep[fig.~13 of ][]{Asencio_2022}, $r_{h, \mathrm{3D}}$ should have been at most $1.1 \times$ smaller at pericentre. This does not significantly affect the conclusions that we draw from the results. Although dwarfs which are only mildly affected by tides will recover their original $r_{h, \mathrm{3D}}$ by the time they reach apocentre, dwarfs which are very affected by tides will only be able to reduce their enhanced $r_{h, \mathrm{3D}}$ values by a small fraction (if they are still bound). Because of this, we consider that the currently observed $r_{h, \mathrm{3D}}$ of Draco and Ursa Minor could also be similar to their last pericentric $r_{h, \mathrm{3D}}$ value, given the high $\eta_{\rm MOND}$ of these dwarfs.

We also note that both in the MOND and in the CDM models, the definition of the tidal radius is an approximation of the limiting Roche Lobe radius through which a dwarf galaxy can be tidally stripped of its mass. The Roche Lobe region is actually triaxial, which means that there are three different tidal radii along the surface of the dwarf. \citet{Baumgardt_2003} chose the first axis ($r_1$) to define $r_{\textrm{tid, Newton}}$ (Eq.~\ref{rtid_LCDM}), while \citet{Asencio_2022} chose the second axis ($r_2$) to define $r_{\textrm{tid, MOND}}$ (Eq.~\ref{rtid_MOND}). Both approximations should, in principle, be valid for describing the observations, but the differences between them should be taken into account when comparing the MOND and the CDM (or Newtonian TDG) results. That is, $r_{\textrm{tid, CDM}}$ and $r_{\textrm{tid, TDG}}$ should be divided by a factor of $3/2$ in order to be directly compared with $r_{\textrm{tid, MOND}}$ \citep{Zhao_2006}.

\subsection{The tidal radius and the tidal susceptibility of the MW satellite galaxies}
\label{rtid_eta_sat}
Using the MW and satellite parameters described in Section~\ref{Classical_sat} and the $r_{\textrm{tid}}$ and $\eta$ equations described in Section~\ref{rtid_eta_theo}, we obtained the $r_{\textrm{tid}}$ and $\eta$ values of the MW satellite galaxies both in the CDM framework (Table~\ref{rtid_eta_LCDM}) and in the MOND framework (Table~\ref{rtid_eta_MOND}).

Additionally, we also indicate the observationally inferred King tidal radius ($r_K$), the radius at which the density profile of a galaxy experiences a sharp drop \citep{King_1966}. Although $r_K$ does not necessarily have to match $r_{\textrm{tid}}$ (e.g. if the satellite galaxy has not experienced any tidal perturbation from the MW), it can provide a constraint on the approximate $r_{\textrm{tid}}$ of a dwarf that has been mildly affected by tides. On the other hand, if the dwarf has been severely affected by tides $-$ and, in the MOND case, if it experiences a significant change in its gravitational regime throughout its orbit \citep{Brada_2000_tides} $-$ it is expected to expand in size after its pericentric passage \citep{Asencio_2022}, making its observed $r_K$ larger than its theoretical $r_{\textrm{tid}}$. Other factors that could cause discrepancies between the theoretical and the observed tidal radius are observational uncertainties (e.g. if observers classify unbound stars as being part of the dwarf) and simplifications in the definition of $r_{\textrm{tid}}$ \citep[see section 2 of][]{van_den_Bosch_2018}.

In the table corresponding to the CDM parameters (Table~\ref{rtid_eta_LCDM}), we also include information on the total mass of the dwarf for this model ($M_{\textrm{tot, CDM}}$). This is obtained by adding $M_{\star}$ and the $M_{DM}$ obtained with Eq~\ref{DM_frac}. In the table corresponding to MOND (Table~\ref{rtid_eta_MOND}), we add a column to indicate how dominant is the baryonic Newtonian gravity of the MW ($g_{_{N, MW}}$) over the baryonic Newtonian self-gravity of the satellite ($g_{_{N, \textrm{dwarf}}} = G M_{\star}/(2 (r_{h, \mathrm{3D}})^2)$) by obtaining the ratio between these two quantities ($g_{\textrm{ratio}} \equiv g_{_{N, MW}}/g_{_{N, \textrm{dwarf}}}$) at pericentre. 

\begin{table*}
	\caption{Properties of the classical satellite galaxies related to tidal effects at pericentre in the CDM model.}
	\centering
	\begin{tabular}{c|c|c|c|c}
		\hline
		Name     & $M_{\textrm{tot, CDM}}$ ($10^6 M_{\odot}$)    & $r_{\textrm{tid, CDM}}$ (pc)    &    $r_K$ (pc)    &  $\eta_{\textrm{CDM}}$   \\ \hline
		\multicolumn{1}{c|}{Fornax}     & $481.32^{+38.99}_{-36.06}$ & $5604.18^{+1787.97}_{-1246.47}$ & $3318.59 \pm 189.38^a$ & $0.17^{+0.04}_{-0.03}$  \\
		\multicolumn{1}{c|}{Carina}     & $51.14^{+1.44}_{-1.40}$    & $3388.56^{+227.32}_{-386.10}$ & $884.68 \pm 45.24^b$  &  $0.12^{+0.01}_{-0.00}$ \\
		\multicolumn{1}{c|}{Draco}      & $34.26^{+0.97}_{-0.94}$    & $1621.78^{+154.50}_{-178.17}$ & $884.19 \pm 62.98^c$ & $0.17^{+0.01}_{-0.01}$ \\
		\multicolumn{1}{c|}{Ursa Minor} & $40.94^{+1.15}_{-1.12}$    & $1723.43^{+129.58}_{-120.45}$ & $753.64 \pm 41.54^d$ & $0.30^{+0.02}_{-0.01}$ \\
		\multicolumn{1}{c|}{Leo~I}      & $189.04^{+31.86}_{-27.27}$ & $3143.68^{+192.72}_{-178.27}$ & $1006.44 \pm 37.03^e$ & $0.11^{+0.01}_{-0.01}$ \\
		\multicolumn{1}{c|}{Leo~II}     & $60.77^{+1.37}_{-1.34}$    & $2850.86^{+365.22}_{-363.46}$ & $632.36 \pm 40.71^f$ & $0.08^{+0.01}_{-0.01}$  \\
		\multicolumn{1}{c|}{Sculptor}   & $110.82^{+8.98}_{-8.30}$   & $2886.44^{+120.15}_{-123.67}$ & $1867.10 \pm 33.38^g$ & $0.12^{+0.01}_{-0.00}$  \\
		\multicolumn{1}{c|}{Sextans}    & $34.45^{+1.17}_{-1.13}$    & $2469.03^{+92.47}_{-91.64}$ & $2238.78 \pm 60.51^h$ & $0.23^{+0.00}_{-0.00}$  \\ \hline
	\end{tabular}
	\tablefoot{The term $r_K$ represents an observable quantity obtained from the angular King tidal radius reported in the following studies: (a) \citet{Wang_2019}, (b) \citet{Irwin_1995}, (c) \citet{Segall_2007}, (d) \citet{Kleyna_1998}, (e) \citet{Sohn_2007}, (f) \citet{Coleman_2007}, (g) \citet{Mateo_1998}, (h) \citet{Roderick_2016}. The angular radius is converted to a physical radius assuming the nominal Galactocentric distance used in this study \citep{Pace_2022}. The errors reported for $r_K$ correspond to the uncertainty in the Galactocentric distance of the dwarf (as this is generally more significant than the uncertainty in the angular radius).}
	\label{rtid_eta_LCDM}
\end{table*}

\begin{table*}
	\caption{Properties of the classical satellite galaxies related to tidal effects at pericentre in MOND.}
	\centering
	\begin{tabular}{c|c|c|c|c}
		\hline
		Name     & $g_{\textrm{ratio}}$   & $r_{\textrm{tid, MOND}}$ (pc)    &    $r_K$ (pc)    &  $\eta_{\textrm{MOND}}$   \\ \hline
		\multicolumn{1}{c|}{Fornax}     & $0.31^{+0.44}_{-0.00}$ & $2455.01^{+1000.27}_{-740.50}$ & $3318.59 \pm 189.38$ & $0.38^{+0.13}_{-0.11}$  \\
		\multicolumn{1}{c|}{Carina}     & $1.87^{+0.62}_{-0.33}$    & $1004.00^{+75.78}_{-105.75}$ & $884.68 \pm 45.24$  &  $0.40^{+0.05}_{-0.02}$ \\
		\multicolumn{1}{c|}{Draco}      & $9.48^{+2.39}_{-2.10}$    & $339.44^{+40.30}_{-35.53}$ & $884.19 \pm 62.98$ & $0.80^{+0.09}_{-0.07}$ \\
		\multicolumn{1}{c|}{Ursa Minor} & $25.23^{+4.80}_{-3.98}$    & $380.64^{+33.56}_{-27.24}$ & $753.64 \pm 41.54$ & $1.36^{+0.11}_{-0.09}$ \\
		\multicolumn{1}{c|}{Leo~I}      & $0.70^{+0.13}_{-0.11}$ & $ 1015.15^{+87.88}_{-90.69}$ & $1006.44 \pm 37.03$ & $0.35^{+0.03}_{-0.03}$ \\
		\multicolumn{1}{c|}{Leo~II}     & $0.75^{+0.28}_{-0.24}$    & $795.79^{+132.61}_{-113.40}$ & $632.36 \pm 40.71$ & $0.27^{+0.04}_{-0.04}$  \\
		\multicolumn{1}{c|}{Sculptor}   & $1.32^{+0.17}_{-0.13}$   & $854.17^{+49.04}_{-44.29}$ & $1867.10 \pm 33.38$ & $0.41^{+0.03}_{-0.02}$  \\
		\multicolumn{1}{c|}{Sextans}    & $12.45^{+0.99}_{-0.92}$    & $627.00^{+24.37}_{-23.52}$ & $2238.78 \pm 60.51$ & $0.92^{+0.03}_{-0.02}$  \\ \hline
	\end{tabular}
	\tablefoot{Since $r_K$ is an observable quantity, it remains the same for both the CDM and the MOND models (see Table~\ref{rtid_eta_LCDM}).}
	\label{rtid_eta_MOND}
\end{table*}

\section{Tidal features in $N$-body simulations}
\label{N-body}
The results of our analysis show that the satellite galaxies are expected to be significantly more affected by Galactic tides in MOND than in the CDM model, where the tidal influence is mostly negligible. This is consistent with several previous works which also found that the classical dwarfs are not tidally susceptible in the CDM framework \citep{Penarrubia_2008, Penarrubia_2009, Battaglia_2015, Iorio_2019, Tokiwa_2023, Jensen_2024, Julio_2025, Boyea_2026}. In order to assess which of the classical satellites' properties can be attributed to tidal effects, and whether this is consistent with their theoretically inferred $\eta$ values, we review the results of $N$-body simulation studies that evaluated tidal effects in dwarf galaxies.

\subsection{$N$-body simulations in the MOND model}
The first MONDian $N$-body simulations that analysed the effect of tides on dwarf galaxies were performed by \citet{Brada_2000_tides}. These were followed by several others \citep{Candlish_2018, Asencio_2022, Bilek_2025}. Since \citet{Asencio_2022} presented their results in terms of the dwarfs' $\eta_{\textrm{MOND}}$ values $-$ as in this study $-$ we mainly focus on their simulations for our analysis. The results of their simulations are concisely displayed in their fig.~13. For an easier comparison, we have reproduced their fig.~13 in Appendix~\ref{MOND_Nbody_sim}.

This figure shows the evolution of a dwarf galaxy's properties throughout its orbit around a spherical gravitational potential. In particular, it shows the evolution of its $r_{h, \mathrm{3D}}$, $\sigma_v$, and 3D aspect ratio (related to $1 - \epsilon$) for different orbital eccentricity ($e$) and $\eta_{\textrm{MOND}}$ values measured at pericentre. Although the central mass and satellites simulated in \citet{Asencio_2022} are more massive than the MW potential and the dwarf satellites, their results regarding the relation between $\eta_{\textrm{MOND}}$ and the tidal disturbance features should still be representative because $\eta$ is a dimensionless quantity. The main parameters that can affect this relation to some extent are $e$, $g_{\textrm{ratio}}$, and $t_{\textrm{peri}}$ \citep{Brada_2000_tides, Asencio_2022}, which we take into account when discussing the results of our analysis. We also note that the relation between $\eta_{\textrm{MOND}}$ and the tidal effects obtained in \citet{Asencio_2022} was calibrated for the same $r_{\textrm{tid, MOND}}$ used in Eq.~\ref{rtid_MOND}, also at pericentre (for the nominal results shown in their fig.~13). The MONDian formulation used in \citet{Asencio_2022} is also the same one as described in our Section~\ref{Classical_sat}.

In the following subsections, we describe the expectations of a MONDian simulation for the kinematic and structural parameters of a tidally perturbed dwarf galaxy. Since the \citet{Asencio_2022} simulations do not investigate the effects of tides on the density profiles of the dwarfs, we also use the results of the MONDian $N$-body simulation of \citet{Bilek_2025} (performed for the Fornax classical satellite) in order to discuss this aspect of tidal disturbance.

\subsubsection{Pericentric enhancement of the radius}
\label{r_h_simulations}
In the $N$-body simulations of \citet{Asencio_2022}, the $r_{h, 3D}$ of several of the simulated dwarfs present an oscillating behaviour, expanding shortly after pericentre and contracting again towards apocentre. In the MOND framework, there are two main reasons for this pericentric radius enhancement: (1) When the dwarf is at its pericentre, it experiences the maximum tidal force of its orbit, and this gravitational pull can cause the dwarf to expand. This effect is expected both in MONDian and in Newtonian gravity (see Section~\ref{radius_simulations_CDM}). (2) As the dwarf approaches the gravitational field of the main galaxy towards its pericentre and then moves away towards its apocentre, it can experience different gravitational regimes. The difference in the $g_{\textrm{ratio}}$ of the dwarf throughout its trajectory is what causes its radius to expand and contract at different points of its orbit \citep{Brada_2000_tides, Wu_2013_phase_trans}. This additional effect (the EFE) is expected in MOND but not in Newtonian gravity.

According to \citet{Brada_2000_tides}, the difference in $g_{\textrm{ratio}}$ is the main mechanism responsible for the oscillation of $r_{h, 3D}$ in the MOND model. Assuming that a dwarf is affected to some extent by the EFE at pericentre, it will experience a more acute oscillation for higher $e$ values. The simulations of \citet{Asencio_2022} show that dwarfs with $e \gtrsim 0.3$ already display this oscillating behaviour. They also show that the amplitude of the oscillation $-$ or, in other words, the degree of radial enhancement right after the pericentric passage $-$ will also be highly dependent on the $\eta_{\rm MOND}$ value of the dwarf at pericentre, since this is correlated with the EFE experienced by the dwarf. Therefore, the higher the $\eta_{\rm MOND}$ and the $e$ values of a dwarf, the more its radius will expand right after pericentre.

If a dwarf is too affected by tides, it can become unstable. That is, it will not be able to contract its radius back to its original state as it moves away from pericentre. This also implies that it will not be able to regain its original properties of $\sigma_v$ and aspect ratio (see Sections~\ref{sigma_simulations} and ~\ref{epsilon_simulations}). The simulations of \citet{Asencio_2022} estimate that the tidal susceptibility threshold at which a dwarf becomes unstable is $\eta_{\textrm{MOND}} > 1.0^{+0.0}_{-0.3}$.

It is important to note that dwarfs that do not experience radial oscillations throughout their orbits (e.g. because they are on quasi-circular orbits), but are sufficiently close to the main galaxy, still display a higher degree of radial expansion with respect to more isolated dwarfs. This is because, due to having $e \approx 0$, these dwarfs are not able to leave the region in which they are affected by the EFE of the galaxy. Conversely, the simulations of \citet{Asencio_2022} also show that unstable dwarfs with higher $e$ values will be able to survive for a longer time, since they spend less time close to pericentre (this does not mean that they will be able to become stable again, but they are able to contract back their radii to a larger extent than dwarfs with lower $e$).

\begin{figure}
	\centering
	\includegraphics[width = 0.5\textwidth]{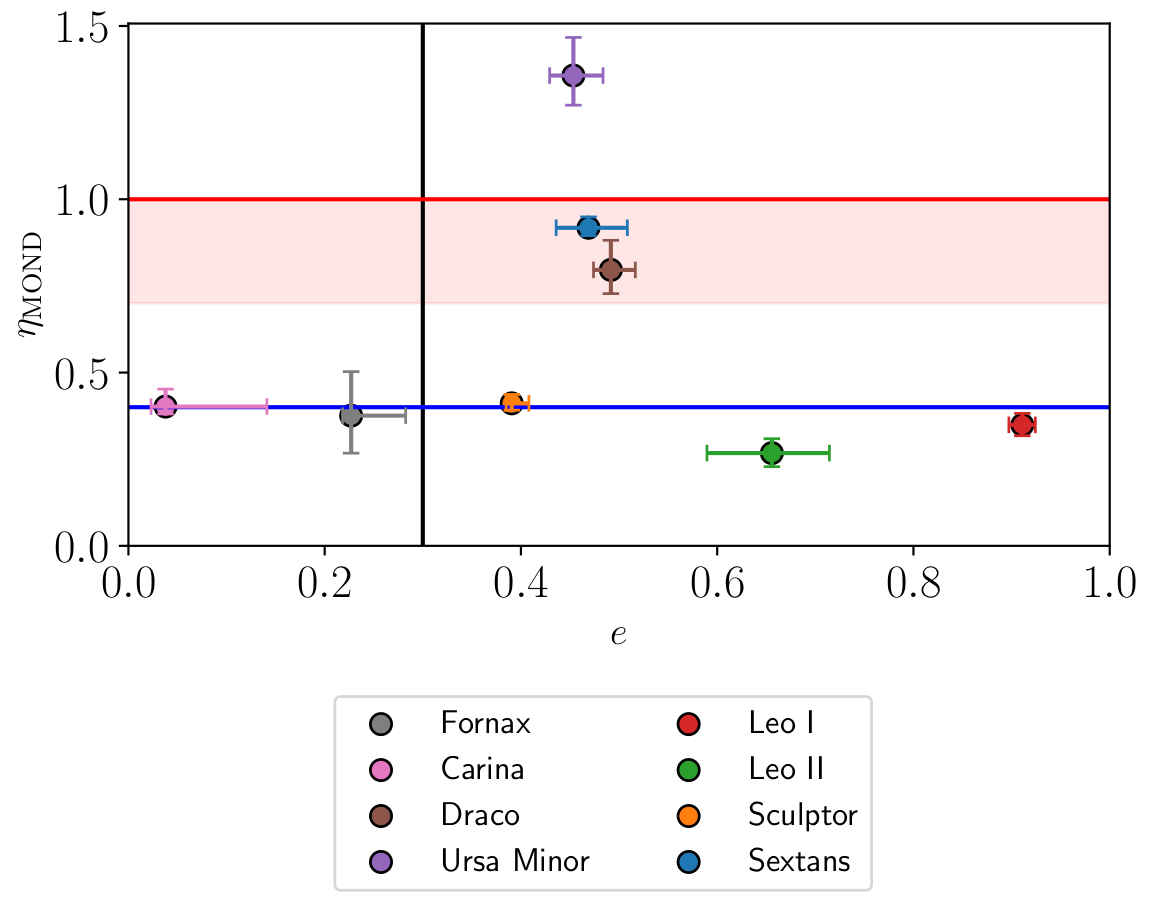}
	\caption{Orbital eccentricity ($e$) vs MONDian tidal susceptibility ($\eta_{\rm MOND}$) of the considered classical dwarfs. The vertical black line represents the approximate $e$ value above which EFE-affected dwarfs are expected to display an oscillating behaviour throughout their orbits. The horizontal blue line represents the approximate $\eta_{\rm MOND}$ value above which dwarfs are expected to experience tidal stripping and an enhancement of their ellipticity. The horizontal red line represents the $\eta_{\rm MOND}$ threshold above which a dwarf becomes unstable (it is not able to contract its radius back to its original size after its pericentric passage), while the shaded area represents the uncertainty in the value of this threshold.}
	\label{e_eta_MOND}
\end{figure}

Fig.~\ref{e_eta_MOND} summarises the predicted properties of the considered classical satellites in the MOND model. According to this, the radius of Draco, Ursa Minor, Leo I, Leo II, Sculptor, and Sextans oscillated throughout their trajectory. Among these, only Ursa Minor, and possibly Draco and Sextans, are in the unstable regime, which means that they are not expected to recover their previous apocentric $r_{h, 3D}$ value in their subsequent apocentric passages. The $r_{h, 3D}$ of Carina and Fornax should not oscillate (or, at least, not as much as in the other dwarfs), but given their proximity to the Galaxy, they could still have their radii enhanced with respect to more distant equivalent dwarfs. The previous constraints can also cause some correlations between the $\eta_{\rm MOND}$ and $e$ values. For instance, the Carina dwarf galaxy cannot escape the region in which it is being tidally perturbed because it has $e \approx 0$. This means that its $r_{h, 3D}$ will be larger than that of an equivalent stable isolated dwarf, and so will its $\eta_{\rm MOND}$. This could explain why Carina and Leo II appear to have similar stellar masses and similar pericentric distances but fairly different $r_{h,3D}$ values. We also note that, although \citet{Asencio_2022} mainly focused on the evolution of $r_{h, \mathrm{3D}}$, the whole dwarf is affected by this radial enhancement and, since the outer layers are more affected by tides, the factor by which the radius is enhanced can be more significant in the outskirts of the dwarf. In other words, the radius enclosing 90\% of the mass might expand to a greater extent.

\subsubsection{Enhancement of the velocity dispersion}
\label{sigma_simulations}
In MOND, dwarfs which are affected by the EFE experience a decrease in their $\sigma_v$ as they approach pericentre due to their radial expansion, but they are able to approximately recover their original $\sigma_v$ at apocentre if they are not affected by tides. Dwarfs which are mildly affected by tides can experience a slight decrease in their $\sigma_v$ across apocentric passages due to mass loss through tidal stripping. An enhancement in $\sigma_v$ is only expected for dwarfs which have $\eta_{\textrm{MOND}} > 1.0_{-0.3}^{+0.0}$. Such dwarfs have experienced tidal disruption so significantly that they cannot recover their original properties after their pericentric passage and have therefore become unstable (i.e, for these dwarfs, the effect of tides is not adiabatic). Therefore, according to this model, only Ursa Minor, and possibly Draco and Sextans, should experience an enhancement of their $\sigma_v$ with respect to tidally unperturbed dwarfs (see Fig.~\ref{e_eta_MOND}).

\subsubsection{Enhancement of the ellipticity}
\label{epsilon_simulations}
The last quantity investigated in the \citet{Asencio_2022} simulations is the aspect ratio of the dwarfs. Their results show that the higher the $\eta_{\textrm{MOND}}$ of the dwarf, the more elliptical the dwarf shape becomes. The squashing effect, i.e. the increase in their ellipticity, can already become appreciable at $\eta_{\textrm{MOND}} \approx 0.4$ and is more significant for dwarfs with higher $e$. Since the lower limit of the $\eta_{\textrm{MOND}}$ threshold at which this effect is expected is not well constrained, any of the considered classical dwarfs could display some degree of flattening, although this should still be less pronounced for dwarfs with $\eta_{\textrm{MOND}} < 0.4$. The same is true regarding the tidal stripping effect (see Section~\ref{density_simulations}).

\subsubsection{Perturbations to the density profile}
\label{density_simulations}
\citet{Bilek_2025} performed MONDian $N$-body simulations of the Fornax dwarf. With these, they found that Fornax-like systems ($\eta_{\textrm{MOND}} \approx 0.4$) are expected to have experienced tidal stripping, so they should present a diffuse stellar halo in their outskirts populated by a group of stars loosely bound to the dwarf. The authors also predict that the Fornax dwarf has a faint stream that has not been detected yet.

We note that although dwarfs that experience some degree of tidal stripping will not be able to recover their exact same properties of $r_{h, 3D}$, $\sigma_v$, and axis ratio, their $r_{h, 3D}$ can still display a quasi-adiabatic behaviour that allows them to approximately recover these properties (the $r_{h, 3D}$ and $\sigma_v$ of tidally stripped, stable dwarfs will decrease slightly across apocentres instead of increasing). Given the long orbital timescales, total disruption by mild tidal stripping might then take longer than a Hubble time. Therefore we still consider these dwarfs to be stable.

Another effect mentioned in \citet{Bilek_2025} is the lopsidedness of the dwarfs' density isophotes, which was also reported by previous studies as a way to distinguish MOND from Newtonian gravity \citep{Wu_2010, Wu_2017, Candlish_2018, Thomas_2018}. However, \citet{Bilek_2025} also mention that, because the deformation of the satellites is expected to take place along the line connecting the satellite and the MW, observers located relatively close to the MW centre (e.g. us) might not be able to detect these deformations. In this regard, observations of external galaxies or clusters would be more helpful.

\subsection{$N$-body simulations in the CDM model}
Several studies have investigated the effect of galactic tides in the CDM model \citep{Kroupa_1997, Mashchenko_2006, Penarrubia_2008, Penarrubia_2009, Penarrubia_2010, Casas_2012, Nichols_2014, Battaglia_2015, van_den_Bosch_2018, Errani_2022, Boyea_2026}. In this paradigm, the tides would first need to strip the outer DM haloes of the dwarfs before they can strip their stellar component \citep{Burkert_1997, Penarrubia_2008,  Battaglia_2015}. If the tidal stripping is sufficiently strong to remove all the outer DM layer, the dwarfs are required to have cuspy DM centres in order to keep their stellar component from becoming completely disrupted after the loss of their outer DM component \citep{Penarrubia_2010, Errani_2022}.

Similarly to MOND, CDM $N$-body simulations also show that tides can only affect the stellar component of a dwarf if its $r_{\textrm{tid, Newton}}$ is comparable to or smaller than its luminous $r_{\textrm{h,3D}}$ \citep{Penarrubia_2009}. Many of the signs of tidal disturbance described in the MOND simulations can also be appreciated in CDM simulations, but with some differences, as we describe below.

\subsubsection{Pericentric enhancement of the radius}
\label{radius_simulations_CDM}
Tidal shocks can also cause the size of the dwarfs to increase with respect to their original radius. However, the degree of radial enhancement is significantly lower than in MOND \citep[see fig. 9d of][]{Mashchenko_2006}. This is because, in Newtonian dynamics, there is no EFE \citep{Brada_2000_tides, Wu_2013_phase_trans}. Despite the tidal shock effect $-$ which may cause the dwarf to puff up at pericentre $-$ if the dwarf experiences significant tidal stripping, its (bound) radius will decrease.

\subsubsection{Enhancement of the velocity dispersion}
Tidally stripped dwarfs generally present a decrease in $\sigma_v$ due to mass loss \citep{Sales_2007}. The gradual mass loss can also produce a velocity gradient along the dwarf \citep{Kroupa_1997, Mayer_2001}. In CDM, an enhancement of $\sigma_v$ only occurs when the dwarf has become unbound. This can only happen if DM cores are assumed to be flat (i.e. constant density) instead of cuspy \citep{Mashchenko_2006}, since cuspy DM profiles keep the dwarf from becoming unbound due to the high DM concentration in the centre. An important difference between the MOND and the CDM models is that, in MOND, it is possible for a dwarf to be unstable (as defined in Section~\ref{r_h_simulations}) and still survive through another pericentric passage because the self-gravity of the dwarf becomes stronger as it moves away from pericentre, allowing the dwarf to recover from the pericentric disruption to some extent. This does not apply to CDM models, where the self-gravity of a dwarf does not vary throughout its orbit.

\subsubsection{Enhancement of the ellipticity}
The CDM simulations show that an increase in the dwarf's $\epsilon$ (and deformations in the dwarf's inner isophotes) can only be appreciated in unbound or very highly disturbed dwarfs \citep[see fig.~12 in][]{Mashchenko_2006}. This is because, if the dwarf presents a cuspy DM centre, this will keep the dwarf from becoming deformed. If it has a flat core, the dwarf can be deformed by the tidal forces, but it will also be extremely vulnerable to tides \citep{Mashchenko_2006, Errani_2022}. We are unlikely to be observing such a dwarf given its short lifetime. \citet{Fattahi_2018} note that some degree of ellipticity enhancement can be appreciated while the dwarf is undergoing tidal stripping, but after $t_{\textrm{cross}}$, this feature should no longer be appreciated and the dwarf should become more spherical than before.

\subsubsection{Perturbations to the density profile}
Tides energize the stars of tidally susceptible dwarfs. Because of this, some stars are able to leave the dwarf through tidal tails or shift onto loosely bound orbits. This effect has been observed in simulations both in CDM \citep{Penarrubia_2009} and in MOND \citep{Bilek_2025} models. By fitting different density profiles to tidally perturbed and unperturbed dwarfs, \citet{Penarrubia_2009} found that tidally perturbed dwarfs are better fitted by a Plummer profile (a power-law profile), while unperturbed dwarfs are better fitted by a King profile (sharp density cut-off in the outskirts).

Other tidal features mentioned in this work are the presence of substructure or irregularities in the density profile. Although these can indeed be caused by tidal effects, the presence of substructure does not necessarily mean that the dwarf has been affected by tides, as other phenomena \citep[e.g. interactions of massive globular clusters; ][]{Bilek_2025} can also generate these features. However, the presence of substructure generally implies that a dwarf is unlikely to have a cuspy DM halo because this will destroy the substructure within $t_{\textrm{cross}}$ \citep{Wilkinson_2005}. While cores can in principle arise in the CDM framework, they are not expected in $\Lambda$CDM when $M_{\star} \la 10^{7.2}~M_{\odot}$ \citep[section~8.1 of][and references therein]{Asencio_2022}.

\section{Comparison with observations}
\label{Observations}
In this section, we compare the expected tidal features of dwarf galaxies with the observational properties of the classical satellites. Since only the MONDian classical satellites are expected to be tidally susceptible, we focus mainly on discussing the observations within this paradigm. We return to the question of how the observed signs of tidal disturbance might be understood in a CDM framework later in the paper (Section~\ref{DM_mass}).

\subsection{Pericentric enhancement of the radius}
The MOND model predicts that EFE susceptible dwarfs with $e \gtrsim 0.3$ will expand their radii shortly after their pericentric passage and contract again towards apocentre, assuming that they are stable ($\eta_{\textrm{MOND}} < 1.0_{-0.3}^{+0.0}$). This enhancement (or the lack of it) can be appreciated in the satellite galaxies by comparing their theoretical $r_{\textrm{tid, MOND}}$ values at pericentre with their presently observed $r_K$ values.

For instance, if a dwarf is not susceptible enough to experience tidal stripping, its $r_{\textrm{tid, MOND}}$ should be larger than its $r_K$. On the other hand, dwarfs which are unstable or currently observed shortly after their pericentric passage are expected to have expanded their radii since pericentre, so $r_K$ should be larger than their pericentric $r_{\textrm{tid}}$. Stable dwarfs which are not currently observed shortly after pericentre with $\eta_{\rm MOND} \gtrsim 0.4$ $-$ the $\eta$ value at which we expect tidal stripping in MOND $-$ should present an $r_{\textrm{tid, MOND}}$ similar to $r_K$.

Based on these predictions, Leo II should present an $r_{\textrm{tid, MOND}}$ larger than its $r_K$, or similar (if the $\eta_{\rm MOND} \gtrsim 0.4$ threshold for tidal stripping adopted in this study was indeed too high). Leo I should have $r_{\textrm{tid, MOND}} \approx r_K$ since it is expected to have already recovered from its pericentric radial expansion, while Carina and Fornax should also exhibit $r_{\textrm{tid, MOND}} \approx r_K$ as they are not expected to have significantly expanded or contracted their radius througout their orbit ($e < 0.3$). Sextans and Sculptor are expected to have $r_K > r_{\textrm{tid, MOND}}$ because they are currently observed shortly after pericentre, while Ursa Minor, and possibly Draco, are expected to have $r_{K} > r_{\textrm{tid, MOND}}$ because they are unstable dwarfs and they should not have been able to recover their pericentric size.

\begin{figure}
	\centering
	\includegraphics[width = 0.5\textwidth]{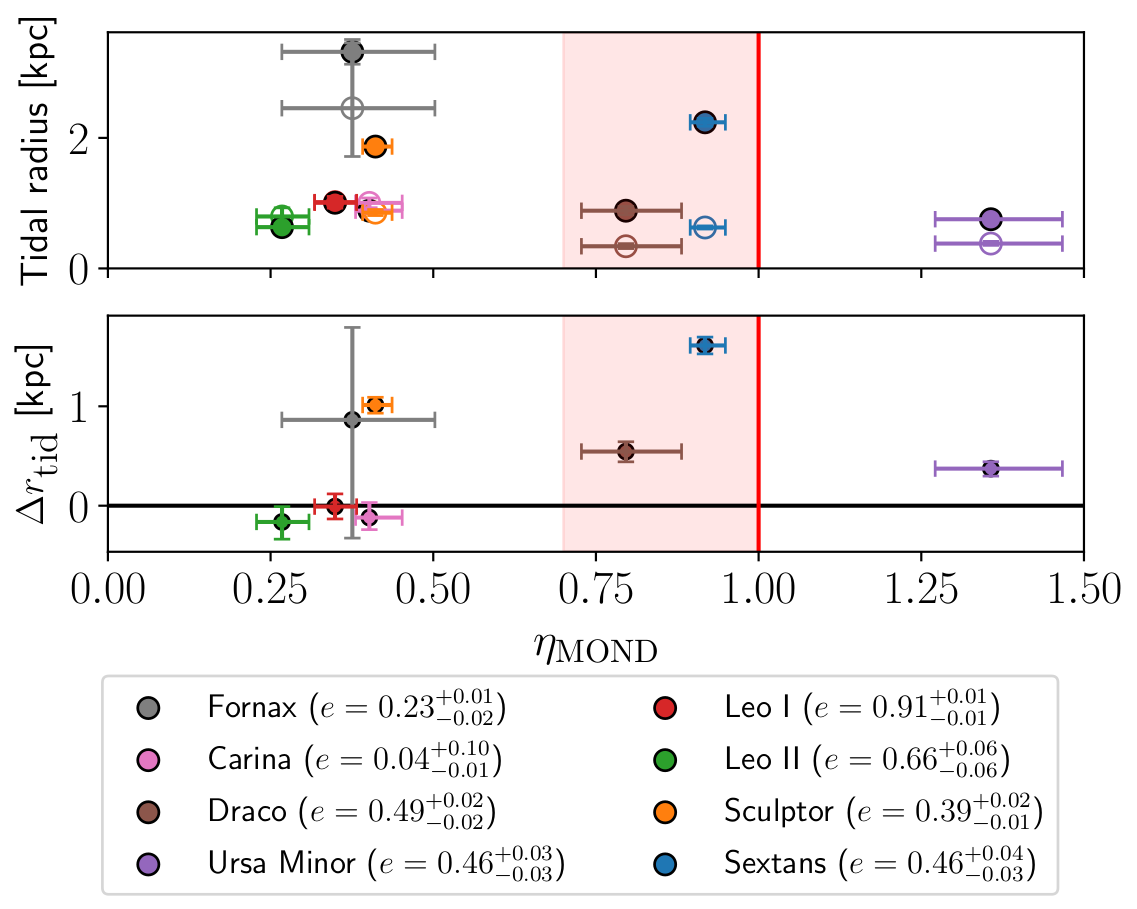}
	\caption{Upper panel: Comparison of the observed King tidal radius $r_K$ (coloured circles) and theoretical MOND tidal radius $r_{\rm tid, MOND}$ at pericentre (empty circles) for the MW classical satellites. Lower panel: Difference between these two quantities $\Delta r_{\rm tid} \equiv r_K - r_{\rm tid, MOND}$. Both figures are plotted against the MOND tidal susceptibility ($\eta_{\rm MOND}$). The red line shows the $\eta_{\rm MOND}$ above which the dwarfs become unstable. The shaded red area accounts for the uncertainty of this threshold.}
	\label{eta_radius_MOND}
\end{figure}

Fig.~\ref{eta_radius_MOND} presents the observed $r_K$ (coloured circles) and the theoretically obtained $r_{\textrm{tid, MOND}}$ at pericentre (empty circles) of the classical satellites vs their $\eta_{\textrm{MOND}}$ values. This shows that the considered sample of classical satellites seems to match the aforementioned expectations within uncertainties. Out of all the considered satellites, only Fornax has a nominal $\Delta r_{\rm tid}$ value that does not approximately match the expectations. However, Fornax also has a very high uncertainty in its $r_{\rm tid, MOND}$ value, which is consistent with the expected $\Delta r_{\rm tid} \approx 0$ within $1\sigma$ confidence. If future observations were to confirm that $\Delta r_{\rm tid} > 0$ for Fornax, it would be necessary to check with detailed $N$-body simulations whether this radial expansion could be caused by other effects (e.g. interactions with Fornax's massive globular clusters).

\subsection{Enhancement of the velocity dispersion}
\begin{figure}
	\centering
	\includegraphics[width = 0.5\textwidth] {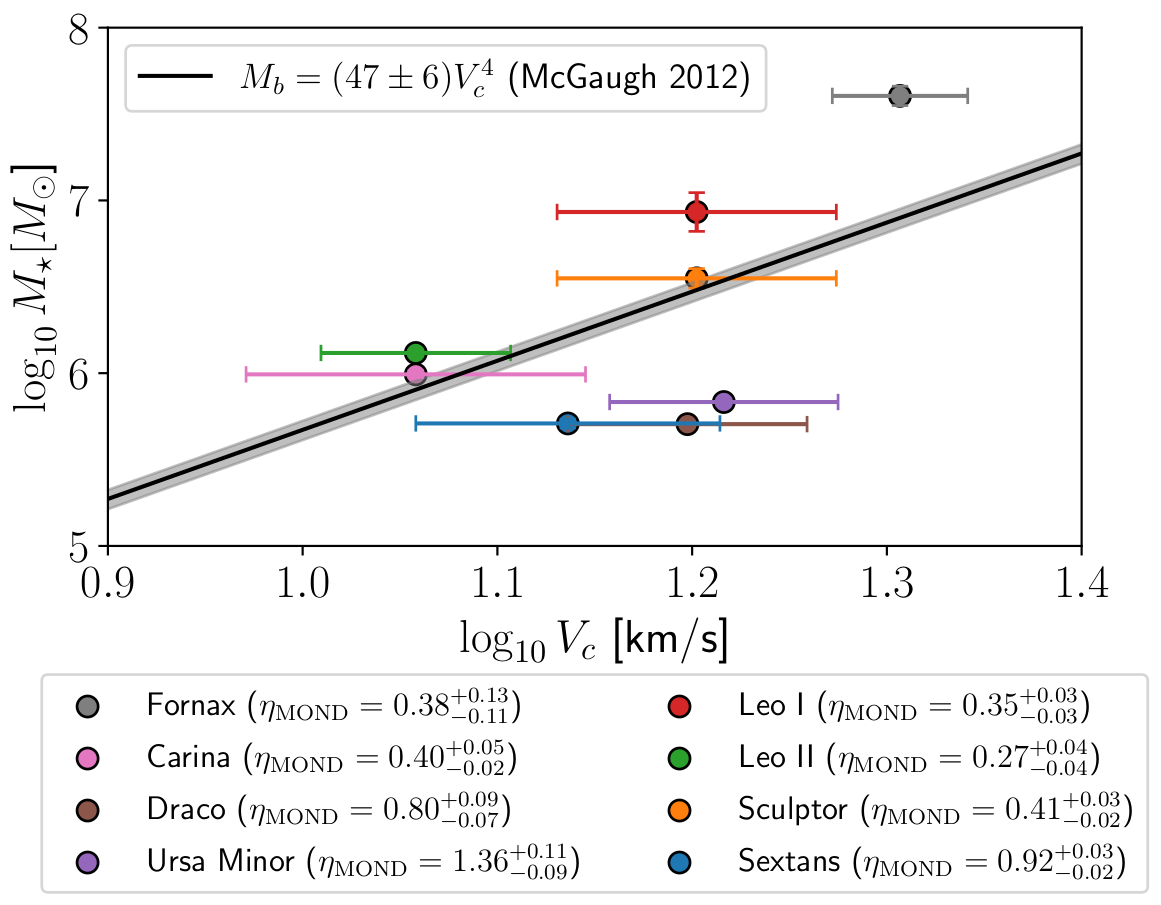}
	\caption{Circular velocity and stellar mass of the classical satellites plotted against the Baryonic Tully-Fisher relation \citep{McGaugh_2012}. The MONDian tidal susceptibility ($\eta_{\textrm{MOND}}$) of each satellite is also provided in the legend.}
	\label{BTFR_MOND}
\end{figure}

In the MOND framework, dwarfs with $\eta_{\textrm{MOND}} > 1.0^{+0.0}_{-0.3}$ should have enhanced $\sigma_v$ values with respect to unperturbed dwarfs, while the $\sigma_v$ of a dwarf with a lower $\eta_{\textrm{MOND}}$ value should be the same (or slightly lower) than that of an unperturbed dwarf. In order to test whether the classical satellites have enhanced $\sigma_v$ values, we estimate their circular velocities ($V_c$) as $V_c = \sigma_v \sqrt{3}$ \citep{Wolf_2010} and, with these, we obtain their positions in the Baryonic Tully-Fisher relation \citep[BTFR;][]{Tully_Fisher_1977}. The BTFR is an empirical relation that shows a tight correlation between the outer flat $V_c$ of isolated galaxies and their $M_{b}$. \citet{McGaugh_2012} fitted this relation as
\begin{eqnarray}
	\frac{M_{b}}{M_{\odot}} = (47 \pm 6) \left(\frac{V_c}{\textrm{km/s}}\right)^4 \, .
\label{BTFR}
\end{eqnarray} 
Fig.~\ref{BTFR_MOND} shows the classical satellites plotted with respect to the BTFR relation (Eq.~\ref{BTFR}). It can be appreciated that the dwarfs with enhanced velocities are also those which present $\eta_{\rm MOND} > 1.0^{+0.0}_{-0.3}$ (Draco, Ursa Minor, and Sextans). We note that since these dwarfs are fairly close to pericentre, they should have had at least another pericentric passage that had already caused their velocity dispersions to increase. We also note that the BTFR relation inferred in \citet{McGaugh_2012} was calibrated for baryonic masses ($M_b = M_{\star} + M_{\rm gas}$) while, in our study, we are only considering the $M_{\star}$ of the dwarfs. Considering the full $M_b$ of the classical satellites would slightly raise their positions in the $y$-axis in Fig.~\ref{BTFR_MOND} but, since the gas content of these dwarfs is significantly smaller than their $M_{\star}$ \citep[$M_{\rm gas} < 10^4~M_{\odot}$;][]{Putman_2021}, we do not expect this to significantly affect our results. The results shown in Fig.~\ref{BTFR_MOND} are consistent with the work of \citet{McGaugh_Wolf_2010}, who also found that deviations from the BTFR are correlated with the dwarfs' susceptibility to tides. These deviations also correlate with the $\epsilon$ of the dwarfs in their study, suggesting that $\epsilon$ could, in turn, be correlated with $\eta$ $-$ as we also propose in our study (see Section~\ref{ellip_obs}).

\begin{figure}
	\centering
	\includegraphics[width = 0.5\textwidth] {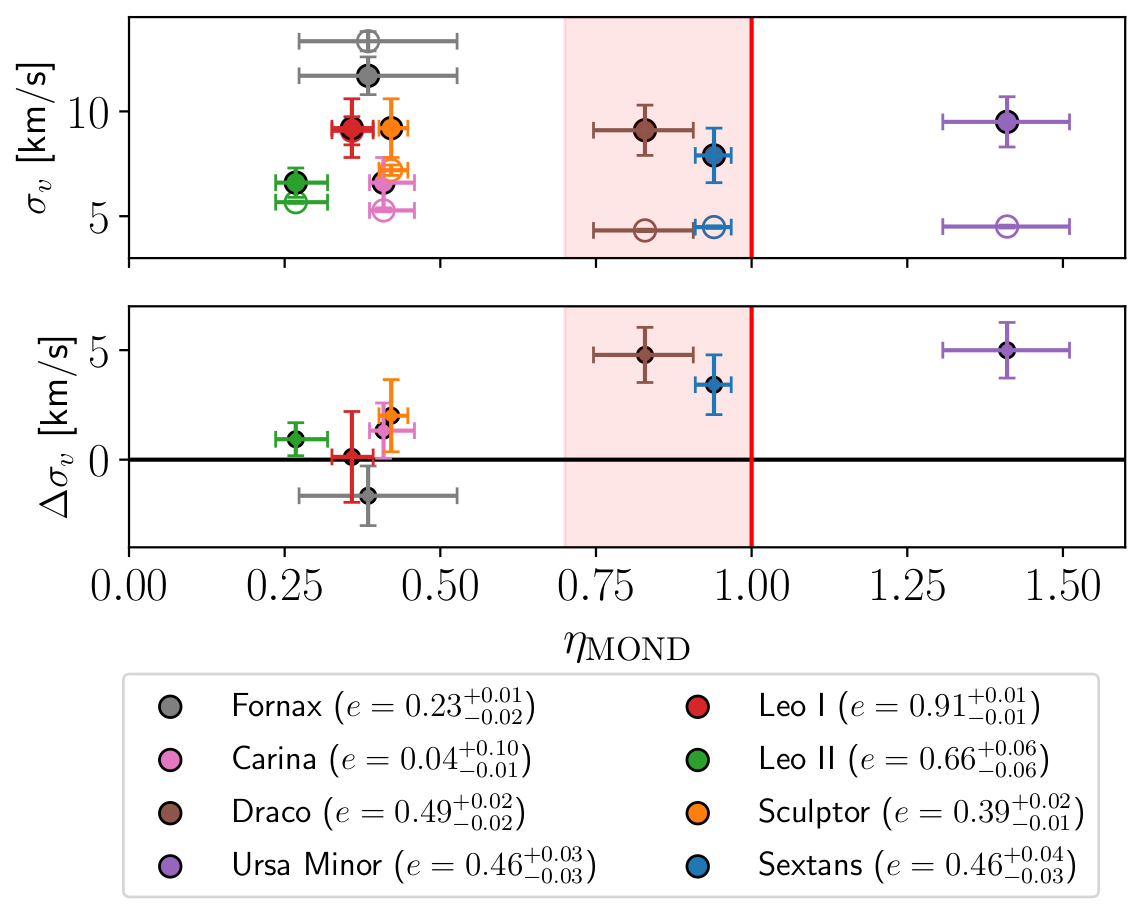}
	\caption{Upper panel: MONDian tidal susceptibility ($\eta_{\textrm{MOND}}$) of the classical satellites vs their velocity dispersion ($\sigma_v$) in observations (filled circles) and in the tidally unperturbed MOND prediction (empty circles). Lower panel: Difference between observations and the tidally unperturbed MOND prediction, which is subtracted. The red line represents the $\eta_{\textrm{MOND}}$ value beyond which an enhancement of $\sigma_v$ is expected with respect to the tidally unperturbed scenario. The shaded red area accounts for the uncertainty in the exact $\eta_{\textrm{MOND}}$ value at which this takes place (see Appendix~\ref{MOND_Nbody_sim}). We note that the theoretical MONDian $\sigma_v$ is obtained at the half-mass radius ($r_{h, \mathrm{3D}}$) of the dwarf, while observations cover a finite aperture centred on the generally higher $\sigma_v$ central region.}
	\label{sigma_MOND}
\end{figure}

An alternative way of testing whether the classical satellites have enhanced $\sigma_v$ values in MOND (due to tides) is by comparing their observed $\sigma_v$ with their theoretically predicted $\sigma_v$ in a tidally unperturbed scenario. To estimate the predicted $\sigma_v (r_{h, 3D})$ in this paradigm, we use eq.~3 of \citet{Haghi_2019_DF2}, which also accounts for the EFE influence on $\sigma_v$. Our results comparing observations (filled circles) to theoretical predictions for a tidally unperturbed scenario (empty circles) are shown in Fig.~\ref{sigma_MOND}.

Similarly to Fig.~\ref{BTFR_MOND}, Fig.~\ref{sigma_MOND} shows that the three satellites with $\eta_{\textrm{MOND}} > 1.0^{+0.0}_{-0.3}$ (Draco, Ursa Minor, and Sextans) are the ones whose $\sigma_v$ is highly enhanced with respect to the tidally unperturbed MOND prediction. The other dwarfs are not expected to have enhanced $\sigma_v$ due to tidal effects, but a few of them (Leo~II, Carina, and Sculptor) seem to have $\sigma_v$ values which are slightly higher than indicated by the theoretical MOND $\sigma_v$. This small discrepancy may not be caused by tidal effects, but by observational limitations. At these low velocity dispersions, results can be inflated by binary stars, though this can be mitigated using multiple observing epochs. Also, the MONDian $\sigma_v$ is obtained at the $r_{h, \mathrm{3D}}$ of the dwarf, while the observed $\sigma_v$ is obtained from many stars at different positions in the dwarf. By looking at the $\sigma_v$ profiles of these dwarfs \citep{Walker_2007, Walker_2009}, it can be appreciated that a significant fraction of the available data is very close to the dwarf's centre. The $\sigma_v$ profile of a galaxy is higher near the centre and softly decreases towards the outskirts, where it becomes flat. Therefore, having many velocity measurements from stars near the centre can lead to a higher $\sigma_v$. Since the $\sigma_v$ profiles of the classical dwarfs are fairly flat, the difference between the $\sigma_v$ values at the centre and at $r_{h, \mathrm{3D}}$ is only $\lesssim 2$~km/s. This difference, even if small, is sufficient to explain the slight $\sigma_v$ discrepancies in Leo~II, Carina, and Sculptor.

It is also important to note that analytical estimates of $\sigma_v$ cannot account for all the effects experienced by the dwarf throughout its orbital history, and that $N$-body simulations might provide more accurate estimates of the $\sigma_v$ values predicted for these dwarfs in MOND. For instance, the $N$-body MONDian simulations of \citet{Bilek_2025} estimate that $\sigma_v \approx 12$~km/s at $r_h$ for Fornax. This value is lower than our analytical estimate of $\sigma_v$ and is in better agreement with observations, even if \citet{Bilek_2025} note that this dwarf's $\sigma_v$ has not been noticeably affected by tides. In this case, the difference with the analytical estimates could be due to memory effects \citep{Kupper_Kroupa_2010, Wu_2013_phase_trans, Haghi_2019_DF2} or to a more detailed reproduction of Fornax's stellar mass profile.

\subsection{Enhancement of the ellipticity}
\label{ellip_obs}
\begin{figure}
	\centering
	\includegraphics[width = 0.5\textwidth] {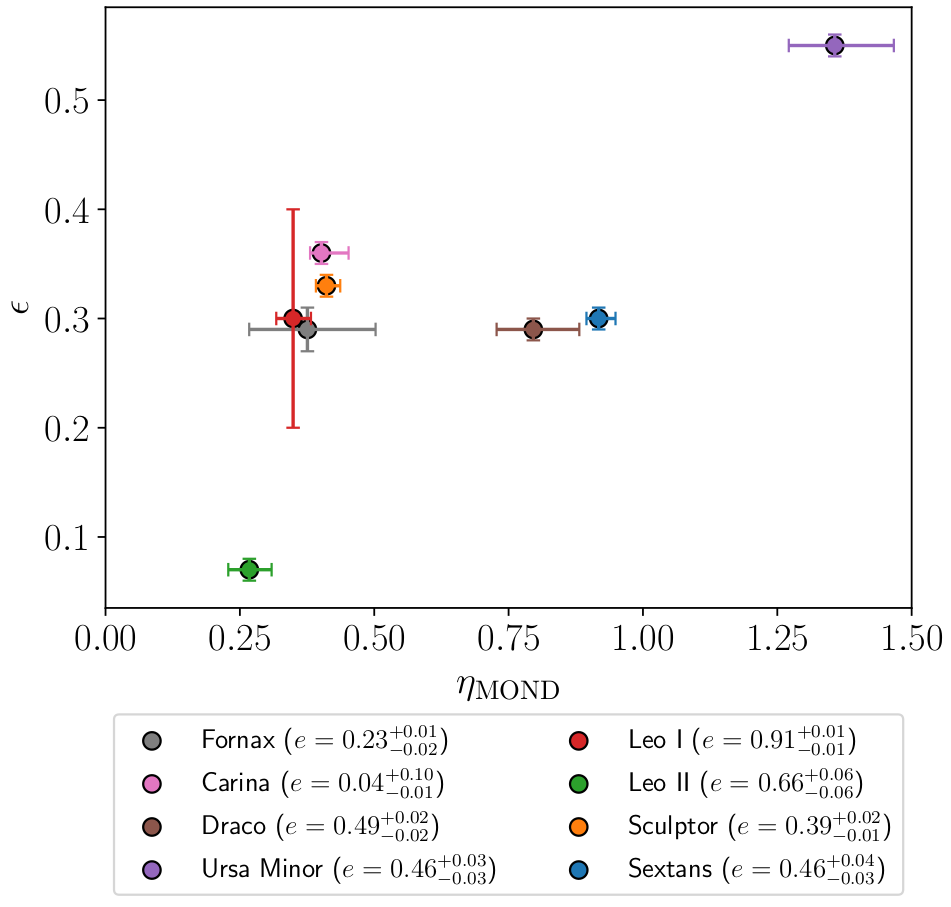}
	\caption{Ellipticity of the classical satellites ($\epsilon$) vs their theoretical tidal susceptibility in MOND ($\eta_{\textrm{MOND}}$). The legend also provides information on the orbital eccentricity ($e$) of these dwarfs.}
	\label{eta_epsilon}
\end{figure}

In MOND, the combination of tidal effects and the EFE should lead to an increase in the $\epsilon$ of the dwarfs. In Fig.~\ref{eta_epsilon}, we show the relation between the $\epsilon$ of the classical satellites and their $\eta_{\textrm{MOND}}$ value. Draco, Ursa Minor, Leo~II, and Sextans seem to match the expected $\eta_{\textrm{MOND}}$ vs $\epsilon$ increasing trend. Fornax and Leo~I could potentially match this trend too, but their uncertainties are currently too large to confirm this. Carina and Sculptor seem to be too elliptical, given their theoretical $\eta_{\textrm{MOND}}$ values. In the case of Carina, this could be explained if the dwarf is currently undergoing tidal stripping, as suggested by the results of \citet{Battaglia_2012}, since the formation of tidal tails can make the dwarf more elliptical \citep{Fattahi_2018}. Also, we note that, although a general increasing trend between $\eta_{\textrm{MOND}}$ and $\epsilon$ is expected, this may not follow a perfectly linear relation due to the fact that the degree of $\epsilon$ enhancement also depends on $e$ and on $t_{\textrm{peri}}$.

\subsection{Perturbations to the density profile}
\label{perturbations_density_profile}
One of the indications that the density profile of a dwarf has been affected by tides is the presence of tidal tails or a stellar overdensity in its outskirts. This tidal effect is expected both in MOND \citep{Bilek_2025} and in CDM \citep{Penarrubia_2009}. By analysing the observed density profiles of the classical MW dwarfs, \citet{Penarrubia_2009} found that Fornax and Leo~II seem to be better fitted by a King profile (no overdensity in the outskirts), while Sagittarius, Leo~I, Sculptor, Carina, Ursa Minor, and Draco are better fitted by a Plummer profile (overdensity in the outskirts). The density profile of Sextans was analysed by \citet{Roderick_2016}, who found that this dwarf's profile was better fitted by a Plummer profile. Subsequent studies \citep{Yang_2022} found that the Fornax dwarf also presents an excess of stars in its outer region. This means that, except for Leo~II, all the analysed classical satellites have density profiles which closely resemble those of tidally perturbed dwarfs. Recent studies \citep{Qi_2022, Sestito_2023, Sestito_2023b, Tolstoy_2025, Ding_2025} have also been able to confirm $-$ using kinematic and metallicity data $-$ that several of the stars beyond $r_K$ in Fornax, Carina, Draco, Sextans, Sculptor, and Ursa Minor are indeed associated with these dwarfs.

From \citet{Bilek_2025}, we can infer that Fornax-like dwarfs ($\eta_{\textrm{MOND}} \approx 0.4$) should already be tidally susceptible enough to experience tidal stripping in MOND. In this paradigm, all the considered classical satellites $-$ except Leo~II $-$ have $\eta_{\textrm{MOND}} \gtrsim 0.4$. This demonstrates good agreement between observations and MOND expectations.

Another potential effect of tides is the formation of substructure and overdensities within the dwarf satellites. Such features have been observed in Fornax \citep{Wang_2019}, Ursa Minor \citep{Irwin_1995, Wilkinson_2005, Munoz_2018, Jensen_2024}, Leo~I \citep{Sohn_2007}, Carina \citep{Lora_2019}, and Sextans \citep{Wilkinson_2005, Roderick_2016, Lora_2019, Cicuendez_2018}. \citet{Lora_2019} also report the presence of substructures in Leo~I and Leo~II, but note that the low number of stars available for these dwarfs makes it difficult to confirm this. Although the presence of substructure can also be attributed to other types of interactions \citep{Bilek_2025}, it strongly suggests that these dwarfs do not have cuspy DM centres \citep{Wilkinson_2005}, in agreement with observations based on dwarf galaxy gas dynamics \citep{Lelli_2022}.

\section{Discussion}
\label{Discussion}
\subsection{Individual MW classical satellites}
\label{Discussion_satellites}
In the previous section, we have shown that the classical satellites present kinematic and morphological features that are consistent with the description of tidally disturbed galaxies. These are expected in a MOND framework, but not in a CDM model where dwarfs are barely affected by tides ($\eta_{\textrm{CDM}} \leq 0.3$). In the following, we discuss the specific properties observed for each of these dwarfs and assess their compatibility with both frameworks. Images of these dwarfs are provided in Appendix~\ref{images_classical_sat} to support the discussion below.
\subsubsection{Fornax}
\label{Fornax}
The Fornax dwarf galaxy is the most massive among the classical satellites of the MW. According to the latest studies \citep{Yang_2022}, it presents an excess of stars in its outer parts. It also presents a fairly elliptical shape \citep[$\epsilon = 0.29 \pm 0.02$; ][]{Munoz_2018} and bursts of star formation which coincide with Fornax's pericentric passages. These are often indicative of the influence of tides. Aside from this, Fornax does not present major signs of being perturbed by tides. Indeed, both the $N$-body simulations of \citet{Bilek_2025} and our analytical $\eta$ estimate indicate that this dwarf should only be mildly tidally disturbed in MOND ($\eta_{\rm MOND} = 0.38^{+0.13}_{-0.11}$). In CDM, this dwarf is expected to be unaffected by tides ($\eta_{\rm CDM} = 0.17^{+0.04}_{-0.03}$). Therefore, this framework requires an alternative explanation that justifies the aforementioned features of this dwarf. \citet{Yang_2022} suggested that Fornax fell into the MW very recently ($<2$~Gyr ago). This would have caused it to lose its gas, consequently leading to a lack of gravity on infall. The residual stars would therefore have spherically expanded to form the observed stellar halo of Fornax. However, in order to explain the correspondence between the star formation episodes of Fornax and its pericentric passages, Fornax must have been orbiting the MW for at least $\approx 5$~Gyr \citep{Bilek_2025}. Those authors showed with MONDian $N$-body simulations that a combination of tides and the EFE could produce a stellar halo very similar to the one reported by \citet{Yang_2022}. If future surveys were to find the faint tidal stream of Fornax predicted by \citet{Bilek_2025}, this would further support that the morphological features of Fornax are influenced by tides.
\subsubsection{Carina}
\label{Carina}
One of Carina's main signs of tidal disturbance is the presence of tidal tails \citep{Battaglia_2012}. This can be explained in MOND, which predicts that the dwarf galaxy is sufficiently affected by tides to experience tidal stripping ($\eta_{\rm MOND} = 0.40^{+0.05}_{-0.02}$), but not in the CDM model, where Carina is not affected by tides ($\eta_{\rm CDM} = 0.12^{+0.01}_{-0.00}$). However, we also note that \citet{McMonigal_2014} found that the Carina tidal tails detected by \citet{Battaglia_2012} were contaminated by a population of stars belonging to the Large Magellanic Cloud. \citet{McMonigal_2014} also pointed out that, with their current data, they could not differentiate between the Magellanic stars and the Carina tidal debris, so the degree of contamination is uncertain. The posterior study of \citet{Qi_2022} was able to confirm, with proper motion information, that several extra-tidal stars are members of Carina, but further data is needed to confirm the presence of tidal tails with this method. If the signal around Carina is indeed a combination of tidal debris and Magellanic stars, instead of a tidal tail structure, the tidal debris around Carina could be attributed to tidal shocks rather than to tidal stripping, as suggested by \citet{Hammer_2020}. Other signs of tidal disturbance in Carina include a fairly elliptical shape \citep[$\epsilon = 0.36 \pm 0.01$;][]{Munoz_2018} and the presence of stellar substructure \citep{Lora_2019}.
\subsubsection{Draco}
Draco is a very interesting object regarding the study of tidal disturbance on dwarf galaxies. On the one hand, it matches almost all the expectations considered in this study of a very tidally disturbed dwarf in MOND $-$ in agreement with $\eta_{\rm MOND} = 0.80^{+0.09}_{-0.07}$. It presents an enhanced radius with respect to its expected pericentric value (see Fig.~\ref{eta_radius_MOND}), an enhanced $\sigma_v$ with respect to tidally unperturbed dwarfs (see Fig.~\ref{sigma_MOND}), a squashed shape \citep[$\epsilon = 0.29 \pm 0.01$;][]{Munoz_2018}, and an overdensity of stars beyond $r_K$ \citep{Penarrubia_2009, Ding_2025}. On the other hand, it presents a very symmetric and smooth density profile, which has led many studies to conclude that Draco is unaffected by tides \citep{Odenkirchen_2001, Piatek_2002, Segall_2007}. We hypothesise that observers could have missed potential signs of morphological distortion if this feature is oriented along the line of sight. With improved data, the line of sight depth of Draco could tell us about its 3D structure and whether it is indeed elongated along the line of sight. Nevertheless, it would be important to check with MOND $N$-body simulations, whether such a smooth profile is expected for Draco in the present day for this framework.

\subsubsection{Ursa Minor}
Ursa Minor is perhaps the classical satellite with the clearest signs of tidal disturbance. It is considered a tidally disturbed dwarf due to morphological asymmetries, large ellipticity \citep[$\epsilon = 0.55 \pm 0.01$;][]{Munoz_2018}, and stellar clumps that are offset from the dwarf’s centroid and appear aligned with Ursa Minor’s orbital direction \citep{Irwin_1995, Palma_2003, Jensen_2024}. It also matches the other expectations of a very tidally disturbed dwarf in MOND, for example, the enhanced radius with regard to its expected pericentric value, the enhanced $\sigma_v$ with regard to tidally unperturbed dwarfs, and an extra-tidal population of stars \citep{Munoz_2005}. This is in good agreement with the high $\eta_{\textrm{MOND}}$ value inferred for this dwarf ($\eta_{\textrm{MOND}} = 1.36^{+0.11}_{-0.09}$). However, $\eta_{\textrm{CDM}} = 0.30^{+0.02}_{-0.01}$ is too small to explain the strong tidal features of Ursa Minor in the CDM paradigm. This was recently confirmed with the Ursa Minor CDM $N$-body simulations of \citet{Boyea_2026}.
\subsubsection{Leo~I}
Leo~I is the most distant of the MW classical satellites, but it is also the classical satellite with the highest orbital eccentricity ($e = 0.91 \pm 0.01$). Because of this and its close pericentric distance ($D_{\textrm{peri}} = 52.44^{+4.67}_{-4.81}$~kpc), it is expected that, in MOND, it might display some signs of tidal disturbance (with $\eta_{\rm MOND} = 0.35^{+0.03}_{-0.03}$). It has indeed been observed to present an excess of stars along the major axis of the main body and an asymmetric radial velocity distribution \citep{Sohn_2007}. Using $N$-body simulations, \citet{Sohn_2007} found that these features are better explained by the effect of tidal disturbance rather than by an extended DM halo around the galaxy. In CDM, Leo~I should not be affected by tides ($\eta_{\textrm{CDM}} = 0.11^{+0.01}_{-0.01}$), so among our two considered paradigms, MOND provides a better explanation for its properties.
\subsubsection{Leo~II}
Leo~II is the dwarf for which we obtained the lowest $\eta$ values in our study, both in MOND ($\eta_{\rm MOND} = 0.27^{+0.04}_{-0.04}$) and in CDM ($\eta_{\textrm{CDM}} = 0.08^{+0.01}_{-0.01}$). It presents a very spherical shape \citep[$\epsilon = 0.07 \pm 0.01$;][]{Munoz_2018}, which is expected for dwarf spheroidal galaxies which have not been very affected by tides. Other studies also report that there are no major signs of tidal disturbance in Leo~II, except for some mild isophotal twisting \citep{Coleman_2007, Koch_2007}. The fact that Leo~II seems to be barely (or not at all) affected by tides is in agreement with expectations both in the MOND and in the CDM model. Still, we note that, in the MOND model, Leo II could have experienced some mild radial enhancement and isophotal squashing at pericentre, but recovered its original shape and size as it moved towards its current position (i.e. close to apocentre). These disturbances may have been what caused the isophotal twisting reported by \citet{Coleman_2007} and \citet{Koch_2007}. New $N$-body simulations of dwarfs with low $\eta_{\rm MOND}$ and high $e$ values are needed to confirm this, since the \citet{Asencio_2022} simulations did not explore this region of the parameter space.
\subsubsection{Sculptor}
The Sculptor dwarf also presents some signs of tidal disturbance, mainly, the presence of a stellar substructure in its outskirts \citep{Westfall_2006, Sestito_2023, Jensen_2024}, a statistically significant radial velocity gradient \citep{Battaglia_2008}, and a fairly high ellipticity \citep[$\epsilon = 0.33 \pm 0.01$;][]{Munoz_2018} that seems to be increasing radially \citep{Jensen_2024}. Within a MOND context, it also presents an enhanced $r_K$ with respect to its pericentric $r_{\textrm{tid, MOND}}$ (see Fig.~\ref{eta_radius_MOND}). These features would, in principle, be better explained by a model in which this dwarf is tidally perturbed, such as MOND. However, we also note that the radial expansion, the $\sigma_v$, and the $\epsilon$ of Sculptor seem to be a bit too high, given Sculptor's $\eta_{\textrm{MOND}} =  0.41^{+0.03}_{-0.02}$, especially when compared with other dwarfs with higher $\eta_{\textrm{MOND}}$ values. These could be explained by observational limitations or by the fact that Sculptor is fairly close to its pericentre (which may enhance these tidal features). It would be necessary to perform MONDian $N$-body simulations of the Sculptor dwarf in order to confirm this. The CDM $N$-body simulations of \citet{Boyea_2026} showed that Sculptor had a close encounter with the Large Magellanic Cloud, which could have further contributed to perturb the dwarf. In the CDM model, this is not sufficient to explain Sculptor's features as tidal, but, in a MOND scenario, it could perhaps explain the slightly too high $r_K$, $\sigma_v$, and $\epsilon$ values of Sculptor.
\subsubsection{Sextans}
Sextans presents several tidal features such as various stellar overdensities and an extended stellar halo in its outskirts \citep{Roderick_2016}. It has also been noted that Sextans' extensive spatial extent could be attributed to the effect of tides \citep{Tokiwa_2023}. This radial enhancement is in fact what MOND predicts for dwarfs on eccentric orbits which are very affected by the EFE, such as Sextans ($e = 0.47^{+0.04}_{-0.03}$, $g_{\rm ratio} = 12.45^{+0.99}_{-0.92}$). The additional tidal features that we described in our study are also well reproduced by Sextans: it displays an enhanced $\sigma_v$ with respect to tidally unperturbed dwarfs and a relatively high ellipticity \citep[$\epsilon = 0.30 \pm 0.01$;][]{Munoz_2018}. Because of this, we also consider Sextans to be in good agreement with MOND expectations.

\subsection{Different DM mass estimates}
\label{DM_mass}
Throughout this work, we have estimated the $\eta_{\textrm{CDM}}$ of the classical satellites by assuming a DM mass fraction given by a sample of dwarf galaxies that are expected to be tidally unperturbed (see Section~\ref{rtid_eta}). The reason for this is that, since tidal effects can potentially affect the $\sigma_v$ of the dwarfs, it can be more reliable to dynamically infer the additional DM mass from systems which are expected to be in dynamical equilibrium.

However, it is, in principle, possible that the classical satellites have a lower DM mass than we assumed. To check this, we obtained the $\eta$ values of these satellites by assuming Newtonian gravity and a mass inferred from their observed $\sigma_v$ values ($M_{\textrm{dyn}}$). The values of $M_{\textrm{dyn}}$, as well as the $r_{\textrm{tid, dyn}}$ and the $\eta_{\textrm{dyn}}$ obtained for this mass, are shown in Table~\ref{M_dyn_Newt_table} for the considered classical satellites. From these, it is possible to see that, even for $M_{\textrm{dyn}}$, the $\eta$ values of the classical dwarfs are still too small in the CDM model to explain their tidal features.

Reducing the DM mass of the classical satellites below their $M_{\textrm{dyn}}$ is not physically sensible for the following reason: in the CDM model, if the enhanced $\sigma_v$ (with respect to purely baryonic Newtonian expectations) of the classical dwarfs is not attributed to DM, it ought to be attributed to tides \citep{Kroupa_1997, Casas_2012}. But, if the dwarfs had lost almost all their DM due to tidal stripping, or were originally DM free TDGs, one would need an additional mechanism to explain the fact that the velocity profiles of these dwarfs appear to be flat \citep{Walker_2007}. Aside from short-lived $\sigma_v$ enhancements at pericentre due to tidal shocks, in a CDM framework, only dwarfs with a cored DM profile can display an enhancement in their $\sigma_v$ due to tidal disturbances, and this enhancement only takes place after they have become unbound or are highly unstable \citep{Mashchenko_2006}. However, the density isophotes of most of the considered classical satellites do not seem to be displaying a tidal destruction process (see, for instance, Fig.~\ref{fig:class_sat_images}).

Other studies have proposed that the presence of substructure and outer stellar haloes in the dwarfs could be due to merger and accretion events \citep{Cicuendez_2018, Jensen_2024, Jensen_thesis, Boyea_2026}. In particular, the simulations of \citet{Deason_2022} and \citet{Goater_2024} showed that late-time dwarf-dwarf mergers can lead to the formation of extended substructure in dwarfs, even without tidal influence. These could perhaps explain the extended stellar haloes of Fornax, Draco, and Sextans in a CDM model, but they are unlikely to explain the tidal tails of the Carina dwarf \citep{Battaglia_2012} or the S-shaped isophotes of Ursa Minor \citep{Palma_2003}. We also note that these mergers are not expected to be very common \citep[$\approx 10\%$ of the MW dwarfs with $M_{\star} > 10^6~M_{\odot}$ are estimated to have undergone a major merger since $z=1$;][]{Deason_2014, Jensen_thesis}, so it would also be a striking coincidence if a large fraction of the most massive ($M_{\star} \gtrsim 10^6~M_{\odot}$) and best observed MW dwarfs had undergone such a merger.

\begin{table*}
	\caption{Tidal radius and tidal susceptibility of the classical satellites obtained for $M_{\textrm{dyn}}$ and for a TDG model at pericentre.}
	\centering
	\setlength{\tabcolsep}{4pt}
	\begin{tabular}{c|c|c|c|c|c}
		\hline
		Name  & $M_{\rm dyn}$ ($10^6 M_{\odot}$)  & $r_{\rm tid, dyn}$ (pc) & $\eta_{\rm dyn}$ & $r_{\rm tid, TDG}$ (pc) & $\eta_{\rm TDG}$\\ \hline
		\multicolumn{1}{c|}{Fornax}  & $212.35 \pm 32.68$ & $4362.52^{+1206.24}_{-1001.94}$ & $0.23^{+0.06}_{-0.04}$ & $2614.06^{+546.54}_{-641.10}$ & $0.40^{+0.09}_{-0.07}$\\
		\multicolumn{1}{c|}{Carina}  & $24.60 \pm 8.95$  & $2703.02^{+118.56}_{-316.86}$   & $0.16^{+0.01}_{-0.00}$ & $905.87^{+40.47}_{-80.71}$   & $0.46^{+0.03}_{-0.01}$ \\
		\multicolumn{1}{c|}{Draco}  & $32.14 \pm 8.48$   & $1583.71^{+152.86}_{-167.49}$  & $0.18^{+0.01}_{-0.01}$ & $394.09^{+33.38}_{-29.76}$  & $0.69^{+0.04}_{-0.03}$ \\
		\multicolumn{1}{c|}{Ursa Minor} & $66.65 \pm 16.84$ & $2027.18^{+146.42}_{-134.69}$  & $0.26^{+0.01}_{-0.01}$ & $441.56^{+26.52}_{-25.92}$   & $1.18^{+0.07}_{-0.05}$ \\
		\multicolumn{1}{c|}{Leo~I}  & $42.24 \pm 12.86$  & $1921.35^{+110.67}_{-129.07}$  & $0.19^{+0.01}_{-0.01}$ & $1131.24^{+63.481}_{-74.63}$  & $0.32^{+0.02}_{-0.03}$ \\
		\multicolumn{1}{c|}{Leo~II}  & $13.54 \pm 2.88$  & $1717.65^{+223.76}_{-202.28}$  & $0.13^{+0.01}_{-0.01}$ & $793.49^{+93.69}_{-84.65}$   & $0.27^{+0.03}_{-0.03}$ \\ 
		\multicolumn{1}{c|}{Sculptor}   & $42.01 \pm 12.79$  & $2098.80^{+75.88}_{-96.56}$   & $0.17^{+0.01}_{-0.01}$ & $920.05^{+32.01}_{-39.13}$  & $0.39^{+0.02}_{-0.01}$ \\
		\multicolumn{1}{c|}{Sextans}  & $50.44 \pm 16.60$ & $2814.72^{+85.64}_{-106.25}$    & $0.21^{+0.00}_{-0.00}$ & $607.95^{+15.40}_{-16.44}$     & $0.95^{+0.01}_{-0.01}$ \\ \hline
	\end{tabular}
	\label{M_dyn_Newt_table}
\end{table*}

\subsection{Comparison with other studies}
\label{Other_studies}
The results of our analysis (Section~\ref{rtid_eta_sat}) imply that in MOND, the classical satellites are expected to be perturbed by tides, which in some cases can lead to an enhancement of their $\sigma_v$. This result differs from the conclusions reached by \citet{Julio_2025}, who determined that the classical satellites are not affected by tides (neither in the CDM nor in the MOND model) and therefore ascribed the enhanced $\sigma_v$ of the classical dwarfs to DM. In the following, we describe the main differences between our analysis and \citet{Julio_2025}.

In our analysis, we obtained $r_{\textrm{tid, MOND}}$ using Eq.~\ref{rtid_MOND} $-$ which is derived for a MONDian potential $-$ and by inferring the stellar mass from the galaxy's luminosity (see Table~\ref{internal_prop}). In \citet{Julio_2025}, both $r_{\textrm{tid, MOND}}$ and $r_{\textrm{tid, dyn}}$ (used indistinctly) are obtained for a Newtonian potential \citep{Pace_2022}, with $M_{\textrm{dyn}}$ used as the mass of the satellite. While the $M_{\textrm{dyn}}$ of the galaxies can encapsulate the MOND effect to some extent (e.g. the MONDian boost to gravity and the EFE), it is generally more accurate to use a MONDian derived equation for $r_{\textrm{tid, MOND}}$. More importantly, the use of $M_{\textrm{dyn}}$ requires the assumption that the dwarf galaxy is in virial equilibrium. This assumption might not be correct \citep[e.g. if the dwarf is affected by tides;][]{Kroupa_1997} and cannot be made for testing whether the dwarf is tidally susceptible. For instance, if the dwarf is significantly perturbed by tides, its $\sigma_v$ will be enhanced, leading to a high $M_{\textrm{dyn}}$ from which a strong self-gravity can be erroneously inferred for the dwarf. This can then lead to the incorrect conclusion that the dwarf's self-gravity is too strong to be affected by tides. Another difference between our study and \citet{Julio_2025} is that we test the $\eta$ of the dwarfs at pericentre, where the tidal effect of the MW is the strongest, while \citet{Julio_2025} test this at their present-day position. In the case of the classical satellites, this can make a difference in the $\eta_{\rm MOND}$ values of some dwarfs (see Appendix~\ref{rtid_eta_present}). Since $N$-body simulations show that pericentric passages in which $\eta$ is sufficiently high can disturb the dwarfs throughout the rest of their trajectory \citep{Asencio_2022}, we argue that it is generally far more useful to consider $\eta$ at pericentre. For all these reasons, we suggest that our analysis provides more reliable results in terms of the tidally susceptible classification of the classical satellites.

Accounting for the fact that Draco, Sextans, and Ursa Minor are sufficiently affected by tides to experience an enhancement in their $\sigma_v$, one can potentially explain the high internal accelerations inferred by \citet{Julio_2025} for these dwarfs. Among the classical satellites, \citet{Julio_2025} also infer from Leo~II's $\sigma_v$ that this dwarf seems to have an enhanced internal acceleration (with respect to MOND expectations), even though this dwarf's $\sigma_v$ does not seem to be too far off from the MOND expectation in our study (see Figs.~\ref{BTFR_MOND} and \ref{sigma_MOND}). The reason for this is that in our study, we chose a more recent dataset for Leo~II's $\sigma_v$ \citep{McConnachie_2020}, while \citet{Julio_2025} used the \citet{Spencer_2017} dataset. In the more recent dataset, some of the Leo~II members in \citet{Spencer_2017} are identified as interlopers and are removed from the dataset. This reduces the $\sigma_v$ of Leo~II sufficiently to reconcile it with MOND. The other classical satellites which are not expected to have experienced a tidal enhancement to their $\sigma_v$ (Fornax, Carina, Sculptor, and Leo~I) seem to be much closer to the tidally unperturbed MOND expectation.

The other dwarf galaxies discussed in \citet{Julio_2025} are not classical MW satellites, so we have not analysed them in detail. Still, we briefly discuss the possible reasons why their inferred internal accelerations also seem to be enhanced with respect to (tidally unperturbed) MOND expectations. Antila~B is the furthest dwarf considered in \citet{Julio_2025} ($1350 \pm 60$~kpc) and it is part of the NGC~3109 association \citep{Sand_2015}. We hypothesise that, in this case, the proximity to NGC~3109 might be affecting Antila~B's dynamics. Eridanus~II is also relatively far away from the MW centre ($366 \pm 17$~kpc), but in \citet{Julio_2025} it is reported to have a similar $r_h$ and a smaller $r_{\textrm{tid}}$ than that of Leo~I $-$ which, in our study, is moderately perturbed by tides. If Eridanus~II were to have a fairly eccentric orbit with a close pericentric passage, its high internal acceleration could be attributed to the effect of tides. Accurate proper motions and MOND orbital integrations including the EFE will be needed to test this. Grus~I presents an $r_{\textrm{tid}}$ significantly smaller than that of Ursa Minor $-$ a dwarf that is being significantly disrupted by tides $-$ and a similar $r_h$. Because of this, we consider it highly likely that the high internal acceleration inferred for Grus I is due to tidal effects. Leo~T has a fairly large $r_{\textrm{tid}}$ at its current position. However, it has been hypothesised that Leo~T is a backsplash galaxy \citep{Blana_2020, Blana_2024}, which means that it could have been affected by the MW in the past. In MOND, backsplash galaxies from the MW could go out to almost 2~Mpc due to the past flyby with M31 \citep[see figure 5 of][]{Banik_2018_anisotropy}.

An interesting feature pointed out by \citet{Julio_2025} is that the internal acceleration profiles of these dwarf galaxies present a hook-like shape in a radial acceleration relation diagram (see their fig.~3), with some regions of the dwarfs exceeding the tidally unperturbed MOND expectations $-$ even in those dwarfs in which a tidal enhancement of $\sigma_v$ is not expected. Similar features have also been described in \citet{Li_2022} and \citet{Mercado_2024}. These studies attributed the upward-bending `hooks' to cuspy DM profiles that enhance the internal acceleration of the dwarfs in the centre. In our work, we disfavour this interpretation for several reasons (see Sections~\ref{perturbations_density_profile} and ~\ref{DM_mass}). We therefore propose a few different explanations for the hook features. One possibility is that the stars which display these high $\sigma_v$ are actually interlopers or within a binary system that is affecting their dynamics \citep{Pianta_2022}. Another possibility is that these stars are close to stellar remnants which are perturbing their $\sigma_v$. For instance, a fairly massive central black hole has been detected in Leo~I \citep{Bustamante-Rosell_2021, Pascale_2024}, which could be affecting the dynamics of this system. In some of the other classical satellites, there have also been X-ray detections that suggest the presence of remnants \citep{Ramsay_2006, Sonbas_2016, Saeedi_2019}. In general, we note that the presence of such remnants in the dwarf galaxy centres is a lot more plausible in MOND than in CDM. The reason for this is that the mass segregation timescale of massive objects can be reduced by a factor of $\approx 10^4$ in MOND $-$ with respect to Newtonian gravity $-$ in dwarf galaxies \citep{Ciotti_2004}. This means that in MOND, it is possible for massive objects to segregate at the centre of their galaxy within a Hubble time. The centres of the classical satellites could therefore be populated by compact clusters of stellar-mass black holes.

Besides \citet{Julio_2025}, \citet{Fattahi_2018} also reported an incompatibility between MOND predictions and the observed large $\sigma_v$ of some of the Local Group dwarf galaxies. But, similarly to \citet{Julio_2025}, this study also did not take into account the possible effect of tides on the $\sigma_v$ of these dwarfs. Although a proper analysis should be conducted to confirm that tides are indeed responsible for this effect, \citet{Fattahi_2018} noted that the dwarfs which presented this $\sigma_v$ enhancement were mainly ultra-faint dwarf galaxies $-$ which are the most tidally susceptible type of dwarf \citep{McGaugh_2010, McGaugh_2021}. In addition to this, \citet{Splawska_2026} also found that some dwarf galaxies which had been reported to present enhanced velocity dispersions experienced a reduction in their inferred $g$ after a more accurate stellar density profile was used to estimate their dynamical properties.

\section{Summary and conclusions}
\label{Conclusions}
In this work, we have studied the tidal susceptibility of the classical satellites of the MW in the MOND and CDM frameworks. The degree of tidal susceptibility of these dwarfs was quantified through the tidal susceptibility parameter $\eta$ (Eq.~\ref{eta}), which we obtained in both models for all the considered dwarfs. Because of the non-linearity of MOND, the internal gravity of these dwarfs can be affected by the external gravitational field of the Milky Way, which dampens the MONDian boost to gravity in these dwarfs. Therefore, it is expected that in MOND the classical satellites will be more susceptible to tides (higher $\eta$ values) than in the CDM model. This is precisely what we find in our analysis.

According to our analysis, the classical dwarfs are mostly unaffected by tides in the CDM model (Ursa Minor is the only dwarf that can be considered to be mildly affected by tides in this paradigm). In the MOND model, some of them are very affected by tides (Draco, Ursa Minor, and Sextans), while others are mildly affected (Fornax, Carina, Leo~I, and Sculptor). Only one of the considered satellites is practically unaffected by tides (Leo~II).

To assess which model provides a theoretical prediction that better describes the observed properties of these dwarfs, we discussed the main tidal features expected in tidally susceptible dwarfs (the enhancement in the dwarf's radius, $\sigma_v$, and $\epsilon$ as well as the tidal perturbations to their density profile). We then compared these expected features with the observed data. Using the $N$-body simulations of \citet{Asencio_2022} as a reference, we were also able to determine the approximate $e$ and $\eta_{\rm MOND}$ values at which each of the considered tidal features starts to become noticeable. For instance, at $\eta_{\rm MOND} \gtrsim 0.4$, the dwarf's shape becomes more elliptical, with the `squashing' effect becoming more significant for higher $\eta$ values. At $e \gtrsim 0.3$, it is expected that the size of an EFE susceptible dwarf will oscillate throughout its trajectory, reaching its maximum size shortly after pericentre. The $\sigma_v$ parameter, however, is only expected to be tidally enhanced in dwarfs that are very tidally perturbed ($\eta_{\rm MOND} \gtrsim 1$). The $\eta_{\rm MOND}$ at which tidal stripping takes place is not analysed in the \citet{Asencio_2022} study, but from \citet{Bilek_2025} we inferred that this effect should take place at $\eta_{\rm MOND} \gtrsim 0.4$. From the simulations, we also found that having higher $e$ values enhances these features and that tidal effects become more significant when the dwarf is close to pericentre. Taking all of this into account, we found good agreement between the observational features of the classical satellites and the $\eta_{\rm MOND}$ values obtained in this study.

We note that in this study we were mainly aiming to check whether the classical satellites match the general trends and features expected for disturbed dwarf galaxies in MOND by using `generic' MONDian $N$-body dwarf simulations of tidally perturbed dwarfs. In order to check whether the exact degree of enhancement predicted by MOND for the radius, the $\sigma_v$, and the $\epsilon$ of the dwarfs, as well as other morphological features, are in agreement with observations, it would be necessary to perform a dedicated $N$-body simulation study for each of these dwarfs $-$ similar to the \citet{Bilek_2025} study about the Fornax dwarf $-$ that accounts for all the known properties of these dwarfs. Given the favourable results shown by our analysis, we expect that this study will motivate further research on the classical dwarfs in the MOND paradigm.

The results of our study indicate that the MOND framework seems to be more promising than the CDM model at explaining the tidal features of the classical dwarfs, given that the low $\eta_{\textrm{CDM}}$ values indicate that these dwarfs are not sufficiently susceptible to tides in the CDM model. In order to assess this thoroughly, we also considered the possibility that the DM fraction assumed in this study overestimates the DM contained in these dwarfs $-$ thus leading to an overestimation of their resilience to tides. To check for this, we re-calculated $\eta_{\textrm{CDM}}$ while assuming instead a DM fraction given by the dynamics ($\sigma_v$) of these dwarfs. Our results showed that even for $M_{\textrm{dyn}}$, the $\eta$ values are too low to explain the tidal features of the classical dwarfs in CDM. The DM fraction cannot be lowered any further because, in this paradigm, the $\sigma_v$ value of the dwarfs can only be explained either by DM or by tidal effects (if the dwarf is extremely disturbed or destroyed by tides). Except for Ursa Minor, the classical dwarfs do not appear to be close to tidal destruction, so the high $\sigma_v$ values of these dwarfs must be ascribed to DM in this paradigm.

In the last part of this study, we briefly discussed other works that claim incompatibilities between MOND and the Local Group dwarf galaxies due to the fact that some of these present very high $\sigma_v$ values. We argue that these could plausibly be explained by tidal effects, as most of these dwarfs are very tidally susceptible. The reason why dwarf galaxies can present an enhanced $\sigma_v$ and still survive destruction (at least for one more pericentric passage) in MOND but not in the CDM models is because MONDian dwarfs can experience various gravitational regimes throughout their trajectories, which means that even if they are very tidally susceptible at pericentre $-$ where they are very affected by the EFE of the main galaxy $-$ they can regain some stability as they become more isolated (and thus more MONDian) towards the apocentre. This `phase-transition' of the gravitational dynamics of an outwards orbiting satellite has been studied in detail by \citet{Wu_2013_phase_trans}. Finally, we also discussed the possibility that the additional $\sigma_v$ enhancement observed for the stars in the centres of some dwarf galaxies could be attributed to the presence of stellar remnants. This is more likely in a MOND scenario since the segregation timescale is significantly shorter in this paradigm.

In conclusion, our results show that the predicted tidal susceptibility of the classical satellites in MOND agrees well with their observed properties. This is not the case in the CDM models, which predict that these dwarfs are barely affected by tides. This constitutes a serious challenge to the theory, and we therefore hope that our results encourage new research on dwarf galaxies in a MOND context, which presently seems to be the most promising paradigm with which to explain the morphological and kinematical features of the classical satellites as well as their planar alignment in a disk of satellites \citep{Pawlowski_2013_VPOS, Bilek_2021, Banik_2022}.

\begin{acknowledgements} 
EA acknowledges support through a teaching assistantship by the Helmholtz-Institut für Strahlen- und Kernphysik. IB is supported by the Royal Society University Research fellowship 211046. PK acknowledges support through the DAAD-Eastern Europe exchange programme between Bonn and Prague. EA is grateful to Prof. Hongsheng Zhao for useful discussions on the mathematical definition of the tidal radius, to Dr. Michal Bílek for consultations on his Fornax dwarf study, and to Mariana Júlio and Dr. Marcel Pawlowski for useful discussions on the radial acceleration profile of the Milky Way satellites. The authors would like to thank the referee for useful comments that significantly improved this manuscript.
\end{acknowledgements}

\bibliographystyle{aa}
\bibliography{MW_satellites}
\onecolumn
\begin{appendix}

\section{MONDian $N$-body simulations of a tidally disturbed dwarf}
\label{MOND_Nbody_sim}
\begin{figure*}[hbt!]
	\centering
	\includegraphics[width = 0.9\textwidth]{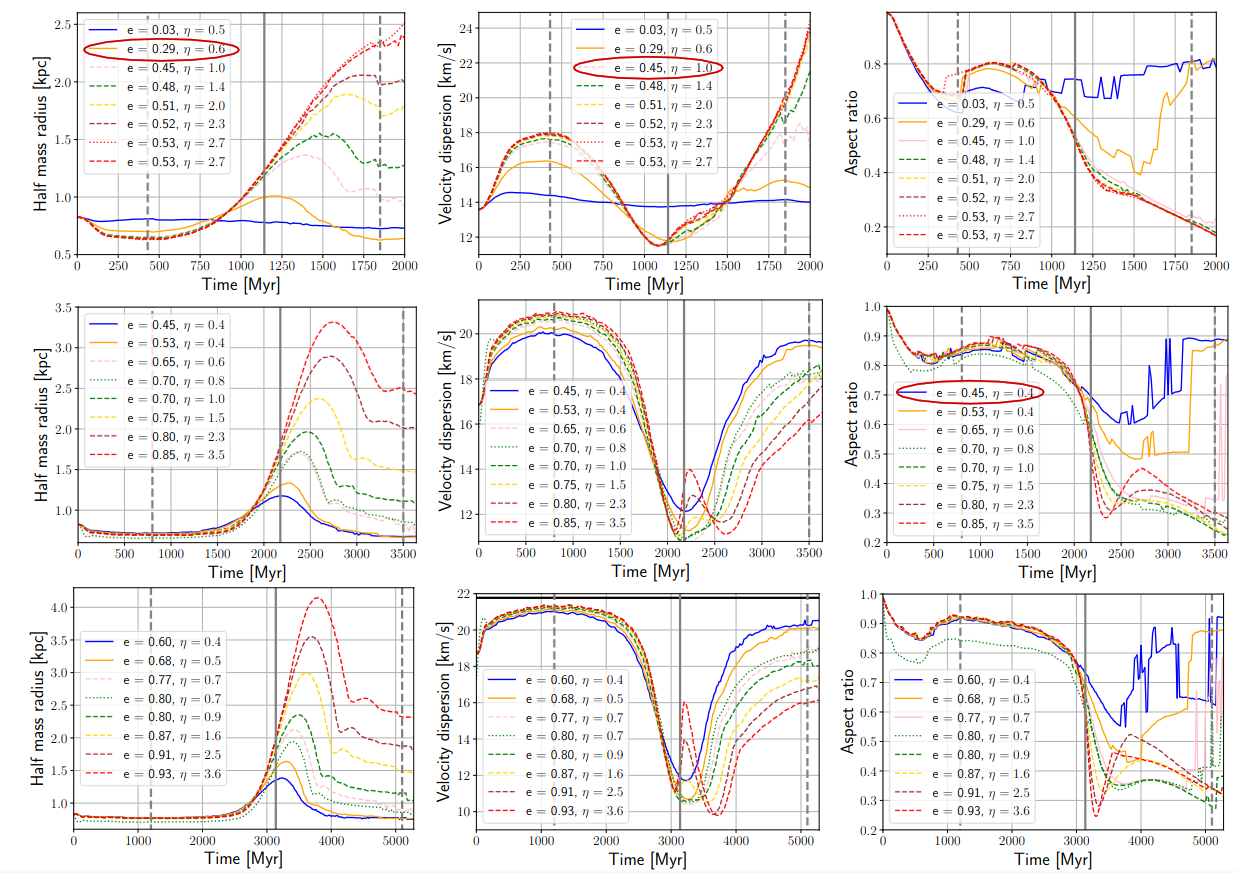}
	\caption{Evolution of $r_{\rm h, 3D}$ (first column), 3D $\sigma_v$ (second column), and aspect ratio (third column) of the MONDian simulated dwarfs over time starting from an initial distance of $R_i = 150$~kpc (first row), $R_i = 300$~kpc (second row), and $R_i = 450$~kpc (third row). The dwarfs are initialised with $r_{\rm h, 3D} = 840$~pc and $M_{\rm dwarf} = 10^{7.5}~M_{\odot}$, orbiting a central potential of mass $M_c = 2.18 \times 10^{12}~M_{\odot}$ at 150~kpc, $M_c = 2.89 \times 10^{12}~M_{\odot}$ at 300 kpc, and $M_c = 3.31 \times 10^{12} M_{\odot}$ at 450~kpc. In each panel, the different curves show different orbital eccentricities as indicated in the legend, which gives their corresponding $\eta_{\rm MOND}$ values at pericentre (solid grey line). The vertical dashed grey lines represent the first and second apocentre of the orbit. The solid (dashed) coloured lines represent those dwarfs which do (do not) recover their initial properties across apocentres. The dotted lines that repeat one of the eccentricities in each panel correspond to a higher resolution simulation, indicating that resolution hardly affects the results. Reproduced from fig.~13 in \citet{Asencio_2022}.}
	\label{fig_Nbody_MOND}
\end{figure*}

Since the results of our MONDian analysis are compared and interpreted in terms of the MONDian $N$-body simulations of \citet{Asencio_2022} for a tidally disturbed dwarf, we provide here (Fig.~\ref{fig_Nbody_MOND}) their fig.~13. In this figure, the authors show the evolution of $r_{h, \mathrm{3D}}$, $\sigma_v$, and the 3D aspect ratio of the tidally disturbed dwarf. The evolution is shown for dwarfs initialised at different initial distances from the central potential and with diferent $e$ values. This allowed us to estimate how the tidal disturbances plus the EFE affect the properties of the dwarf at various $\eta_{\rm MOND}$ values. This also allowed us to compare the degree of disturbance between dwarfs with the same $\eta_{\rm MOND}$ values but with different $e$ values.

In Fig.~\ref{fig_Nbody_MOND}, we have highlighted with a red ellipse the configurations ($\eta_{\rm MOND}$, $e$) that we have used as reference in our analysis in order to constrain the $\eta_{\rm MOND}$ and the $e$ values at which we expect to see certain tidal effects. For instance, for EFE susceptible dwarfs with appreciable orbital eccentricity ($e \gtrsim 0.3$), we expect that their radii will oscillate throughout their trajectory, achieving a maximum shortly after their pericentre. Regarding the $\sigma_v$ enhancement, we find that the value at which the dwarf starts to become unstable $-$ that is, it cannot recover its properties of $r_h$, $\sigma_v$, and aspect ratio across apocentric passages $-$ is $\eta_{\rm MOND} \approx 1.0$. The lower limit of this value is not very well constrained. At $\eta_{\rm MOND} = 0.6$, the dwarfs still seem to be fairly stable, but the parameter space between $\eta_{\rm MOND} = 0.6$ and $\eta_{\rm MOND} = 1.0$ has not been explored. Because of this, we consider that enhancements in the $\sigma_v$ of the dwarfs should be expected beyond $\eta_{\rm MOND} = 1.0^{+0.0}_{-0.3}$. The aspect ratio of the dwarfs is shown to be already disturbed for the lowest $\eta_{\rm MOND}$ values tested in the simulation $-$ $\eta_{\rm MOND} = 0.5$ for the quasi-circular orbits and $\eta_{\rm MOND} = 0.4$ for the more eccentric orbits $-$ but it is plausible that it can also be disturbed for even lower $\eta_{\rm MOND}$ values.

\FloatBarrier

\section{Tidal radius and tidal susceptibility at the present day}
\label{rtid_eta_present}
For our nominal analysis, we have estimated the $r_{\rm tid}$ and the $\eta$ values of the classical dwarfs at pericentre $-$ since this is the position at which they experience the maximum effect from tides. In Table~\ref{rtid_eta_present_nominal}, we present these values for the dwarfs' present day positions in the MOND model and in the nominal CDM model. In Table~\ref{rtid_eta_present_class} we present the values of $r_{\rm tid}$ and $\eta$ for a CDM model in which we infer the DM fraction of the classical satellites from their $\sigma_v$, and a Newtonian model in which we assume the classical satellites to be TDGs and, therefore, free of DM.

\begin{table*}[!htbp]
	\caption{Properties of the classical satellites at their present day location in the MOND and (nominal) CDM models.}
	\centering
	\begin{tabular}{c|c|c|c|c|c}
		\hline
		Name  &  $g_{\textrm{ratio}}$   & $r_{\textrm{tid, MOND}}$ (pc)    &  $\eta_{\textrm{MOND}}$ & $r_{\textrm{tid,}\textrm{CDM}}$ (pc) & $\eta_{\textrm{CDM}}$ \\ \hline
		\multicolumn{1}{c|}{Fornax} & $0.19^{+0.01}_{-0.02}$ & $4729.44^{+228.42}_{-316.28}$ & $0.23^{+0.01}_{-0.01}$ &  $8779.22^{+388.69}_{-640.36}$ & $0.13^{+0.00}_{-0.00}$\\
		\multicolumn{1}{c|}{Carina}  &  $1.95^{+0.15}_{-0.24}$ & $991.89^{+42.63}_{-59.03}$ & $0.41^{+0.01}_{-0.02}$ & $3327.35^{+132.23}_{-217.85}$ & $0.12^{+0.00}_{-0.00}$ \\
		\multicolumn{1}{c|}{Draco} & $3.68^{+0.17}_{-0.65}$ & $562.13^{+33.42}_{-46.96}$ & $0.49^{+0.02}_{-0.02}$ &  $2290.51^{+131.05}_{-237.04}$ & $0.12^{+0.00}_{-0.00}$\\
		\multicolumn{1}{c|}{Ursa Minor} & $9.08^{+0.61}_{-1.10}$ & $639.86^{+29.15}_{-40.36}$ & $0.83^{+0.02}_{-0.03}$ & $2482.55^{+106.02}_{-174.66}$ & $0.21^{+0.00}_{-0.00}$ \\
		\multicolumn{1}{c|}{Leo~I}  & $0.03^{+0.00}_{-0.00}$ & $5080.10^{+75.90}_{-296.41}$ & $0.07^{+0.00}_{-0.00}$ & $9473.89^{+269.55}_{-444.10}$ & $0.04^{+0.00}_{-0.00}$ \\
		\multicolumn{1}{c|}{Leo~II}  &  $0.09^{+0.01}_{-0.01}$ & $2380.61^{+130.72}_{-181.00}$ & $0.09^{+0.00}_{-0.00}$ &  $6004.84^{+299.61}_{-493.60}$ & $0.09^{+0.00}_{-0.00}$ \\
		\multicolumn{1}{c|}{Sculptor}   &  $0.68^{+0.01}_{-0.03}$ & $1196.72^{+24.07}_{-18.95}$ & $0.30^{+0.00}_{-0.01}$ &  $3661.44^{+55.16}_{-84.02}$ & $0.10^{+0.00}_{-0.00}$ \\
		\multicolumn{1}{c|}{Sextans}  & $9.63^{+0.36}_{-0.70}$ & $710.68^{+21.40}_{-16.61}$ & $0.81^{+0.01}_{-0.02}$ & $2706.95^{+56.43}_{-92.98}$ & $0.21^{+0.0}_{-0.00}$ \\ \hline
	\end{tabular}
	\label{rtid_eta_present_nominal}
\end{table*}

\begin{table*}[!htbp]
	\caption{Properties of the classical satellites obtained for $M_{\textrm{dyn}}$ and for the TDG scenario at their present day positions.}
	\centering
	\begin{tabular}{c|c|c|c|c}
		\hline
		Name  & $r_{\rm tid}$ (pc) & $\eta_{\rm dyn}$ & $r_{\rm tid, TDG}$ (pc) & $\eta_{\rm TDG}$\\ \hline
		\multicolumn{1}{c|}{Fornax}  & $6674.17^{+330.35}_{-457.41}$ & $0.17^{+0.00}_{-0.00}$ & $3842.46^{+106.38}_{-175.15}$ & $0.29^{+0.00}_{-0.01}$\\
		\multicolumn{1}{c|}{Carina}  &  $2604.34^{+115.67}_{-160.17}$   & $0.15^{+0.00}_{-0.00}$ & $892.70^{+22.64}_{-40.96}$   & $0.45^{+0.01}_{-0.01}$ \\
		\multicolumn{1}{c|}{Draco}  &  $2237.06^{+140.45}_{-194.48}$  & $0.12^{+0.00}_{-0.00}$ & $565.53^{+19.69}_{-36.97}$  & $0.49^{+0.01}_{-0.02}$ \\
		\multicolumn{1}{c|}{Ursa Minor} &  $2916.69^{+139.22}_{-192.77}$  & $0.18^{+0.00}_{-0.00}$ & $634.22^{+17.22}_{-31.15}$   & $0.83^{+0.02}_{-0.02}$ \\
		\multicolumn{1}{c|}{Leo~I}  &  $5839.88^{+88.64}_{-346.17}$  & $0.06^{+0.00}_{-0.00}$ & $3372.70^{+83.12}_{-79.16}$  & $0.11^{+0.00}_{-0.00}$ \\
		\multicolumn{1}{c|}{Leo~II}  &  $3634.26^{+202.82}_{-280.83}$  & $0.06^{+0.00}_{-0.00}$ & $1670.53^{+71.05}_{-86.07}$   & $0.13^{+0.00}_{-0.00}$ \\
		\multicolumn{1}{c|}{Sculptor}   &  $2635.41^{+55.76}_{-43.28}$   & $0.13^{+0.00}_{-0.00}$ & $1163.71^{+11.99}_{-17.27}$  & $0.31^{+0.00}_{-0.00}$ \\
		\multicolumn{1}{c|}{Sextans}  & $3048.59^{+95.72}_{-74.30}$    & $0.19^{+0.00}_{-0.00}$ & $664.95^{+11.97}_{-12.91}$   & $0.87^{+0.01}_{-0.01}$ \\ \hline
	\end{tabular}
	\label{rtid_eta_present_class}
\end{table*}
\FloatBarrier

\section{Density isophotes of classical satellites}
\label{images_classical_sat}
\begin{figure*}[hbt!]
  \centering
  \begin{subfigure}{.48\textwidth}
    \centering
    \includegraphics[width=1.\linewidth]{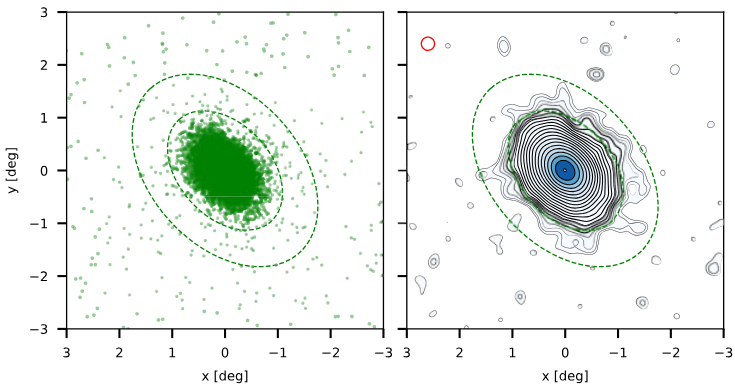}
    \caption{Fornax}
    \label{fig:Fornax}
  \end{subfigure}
  \hfill
  \begin{subfigure}{.48\textwidth}
    \centering
    \includegraphics[width=0.85\linewidth]{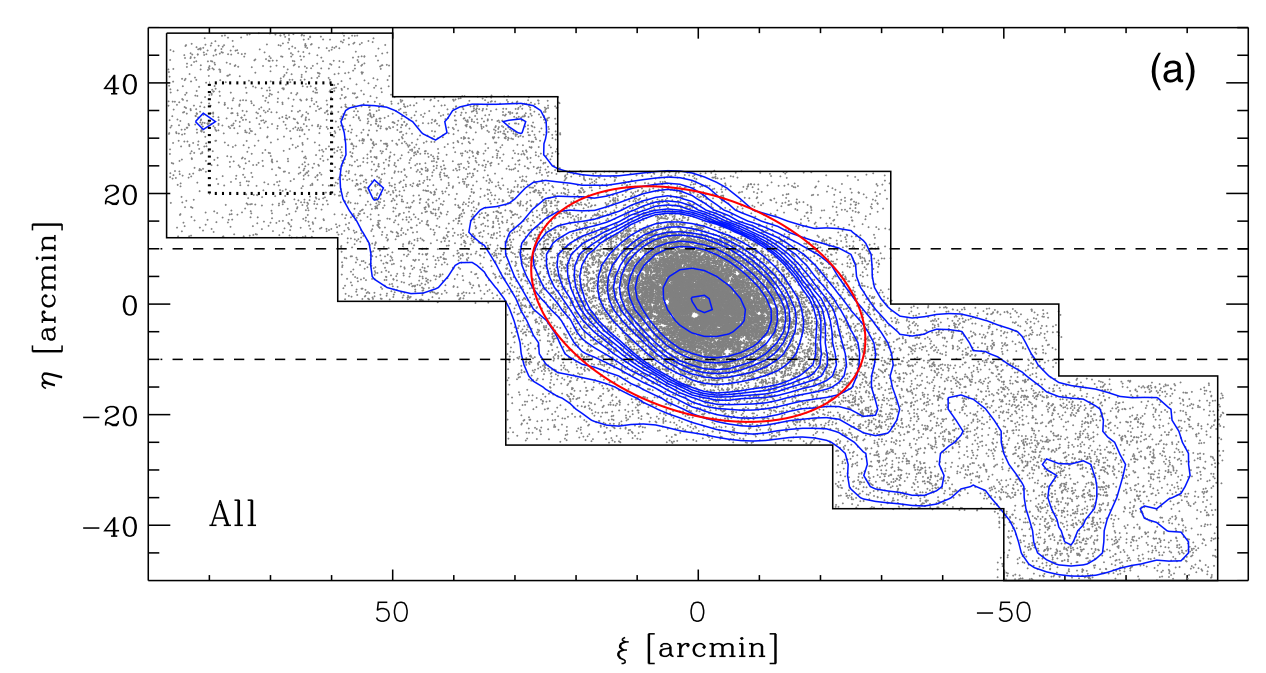}
    \caption{Carina}
    \label{fig:Carina}
  \end{subfigure}

  \vspace{1em}

  \begin{subfigure}{.32\textwidth}
    \centering
    \includegraphics[width=0.9\linewidth]{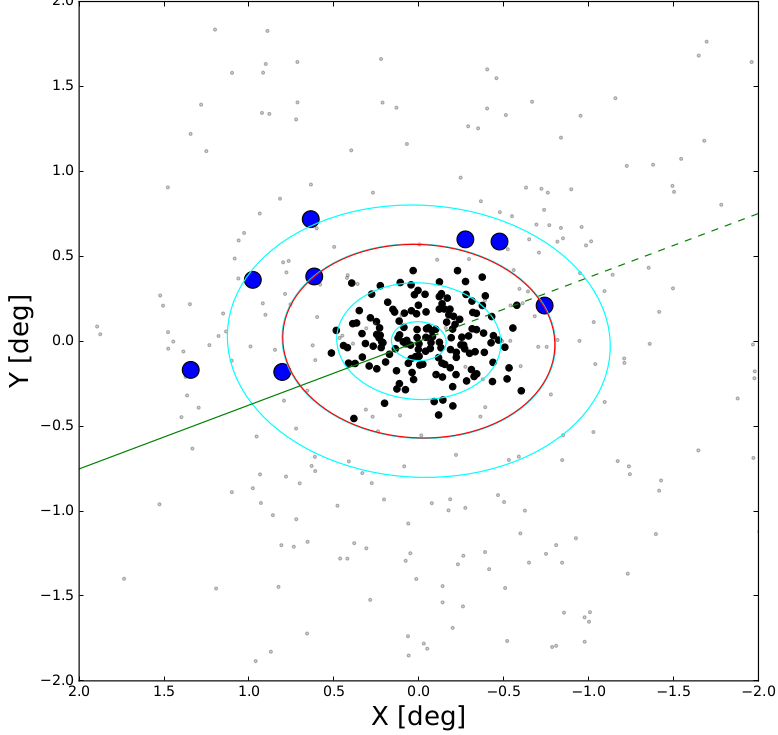}
    \caption{Draco}
    \label{fig:Draco}
  \end{subfigure}%
  \begin{subfigure}{.32\textwidth}
    \centering
    \includegraphics[width=.9\linewidth]{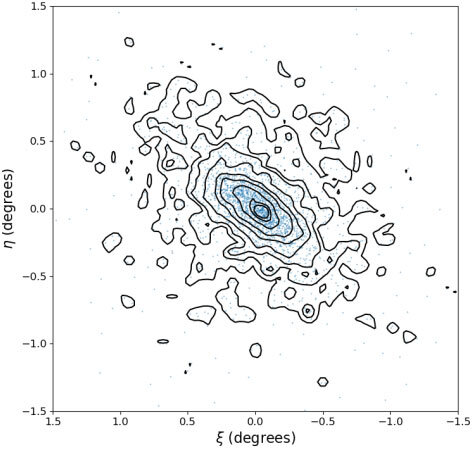}
    \caption{Ursa Minor}
    \label{fig:UrsaMinor}
  \end{subfigure}%
  \begin{subfigure}{.32\textwidth}
    \centering
    \includegraphics[width=.9\linewidth]{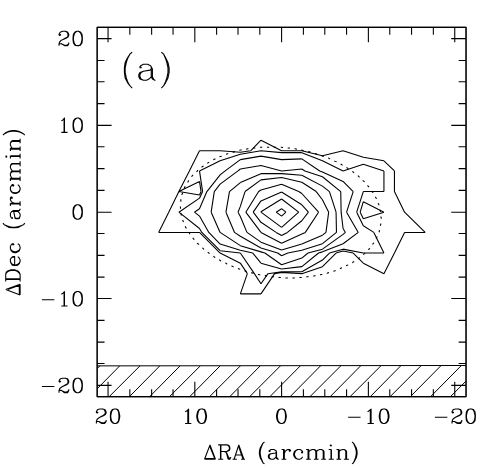}
    \caption{Leo~I}
    \label{fig:LeoI}
  \end{subfigure}

  \vspace{1em}

  \begin{subfigure}{.33\textwidth}
    \centering
    \includegraphics[width=.9\linewidth]{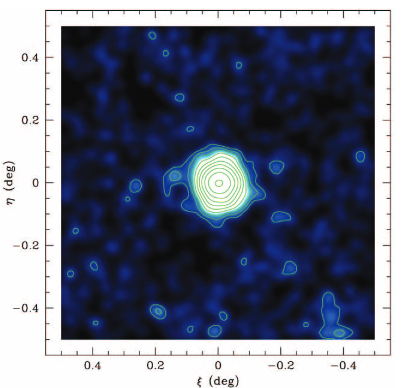}
    \caption{Leo~II}
    \label{fig:LeoII}
  \end{subfigure}%
  \begin{subfigure}{.33\textwidth}
    \centering
    \includegraphics[width=.9\linewidth]{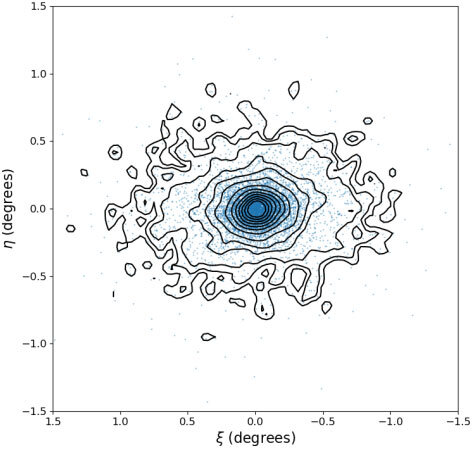}
    \caption{Sculptor}
    \label{fig:Sculptor}
  \end{subfigure}%
  \begin{subfigure}{.33\textwidth}
    \centering
    \includegraphics[width=.95\linewidth]{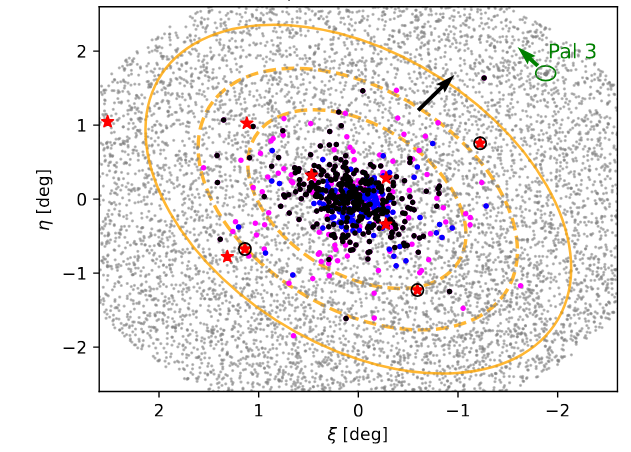}
    \caption{Sextans}
    \label{fig:Sextans}
  \end{subfigure}%

\caption{(a) Stellar distribution (left panel) and contour map (right panel) of the Fornax dwarf. Figure reproduced from fig.~6 in \citet{Yang_2022}. The dashed green circles represent the range within which the outer stellar halo can be appreciated. (b) Spatial distribution of stars in Carina with overlaid isodensity contours. Figure reproduced from fig.~3 in \citet{Battaglia_2012}. The red ellipse corresponds to the nominal $r_K$ assumed in their study. (c) Spatial distribution of the Draco stars. Figure reproduced from fig.~7 in \citet{Ding_2025}. The black dots represent the high probability members of Draco, identified using line of sight velocity and metallicity information. The large blue dots highlight the eight stars outside $r_K$ which have also been confirmed to be high probability members of Draco. Cyan ellipses are plotted at 1, 3, 5, and 7 half light radii, with the red ellipse overlapping with 5 half light radii being the $r_K$ inferred in their study. (d) Isophote contour map of Ursa Minor. Figure reproduced from fig.~11 in \citet{Jensen_2024}. (e) Isodensity contour plot of Leo~I. Figure reproduced from fig.~11 in \citet{Sohn_2007}. The dashed ellipse is their inferred $r_K$. (f) Stellar density contour diagram of Leo~II. Figure reproduced from fig.~9 in \citet{Coleman_2007}. (g) Isophote contour map of Sculptor. Figure reproduced from fig.~10 in \citet{Jensen_2024}. (h) Spatial distribution of Sextans members. Figure reproduced from fig.~9 in \citet{Tolstoy_2025}. The orange lines represent, from outwards to inwards, the $r_K$ inferred in \citet{Irwin_1995}, \citet{Cicuendez_2018}, and \citet{Roderick_2016}. The coloured dots and star symbols represent stars from different catalogues and surveys that have been classified as Sextans members. The black arrow is the mean proper motion of Sextans, while the green arrow is the same for the Palomar 3 globular cluster.}
\label{fig:class_sat_images}
\end{figure*}

For reference, in this section, we display the stellar distribution and the isophote contour maps of the classical satellites, as given by previously published studies. The Fornax dwarf image (Fig.~\ref{fig:Fornax}) displays an elliptical shape and irregularities in its outskirts, which have been identified as an outer stellar halo \citep{Yang_2022}. The Carina dwarf image (Fig.~\ref{fig:Carina}) shows the tidal tails connected to this dwarf, as well as its elliptical shape. The Draco dwarf image (Fig.~\ref{fig:Draco}) shows the presence of several Draco members beyond its $r_K$. The Ursa Minor image (Fig.~\ref{fig:UrsaMinor}) shows that this dwarf is significantly perturbed, with a highly elliptical shape and a dispersed and disjointed structure. The Leo~I image (Fig.~\ref{fig:LeoI}) shows that the dwarf presents a fairly elliptical shape and irregularities in its outskirts, as well as several stars beyond its $r_K$. Leo~II (Fig.~\ref{fig:LeoII}) presents very mild ellipticity. The mild isophotal twisting reported in \citet{Coleman_2007} can also be appreciated in the image. Sculptor (Fig.~\ref{fig:Sculptor}) presents a stellar distribution which becomes progressively more elliptical and irregular towards the outskirts. The Sextans image (Fig.~\ref{fig:Sextans}) shows a very spatially extended and diffuse dwarf with several stars outside its $r_K$, even for the oldest and most conservative $r_K$ estimate \citep{Irwin_1995}.

\label{lastpage}
\end{appendix}
\end{document}